\documentclass[a4paper,10pt,prl,twocolumn,superscriptaddress]{revtex4}
\usepackage{color,amsmath,amssymb,amsthm,times,graphics,graphicx,bm,bbm,dcolumn}

\newcommand{\be}{\begin{equation}}
\newcommand{\ee}{\end{equation}}

\newcommand{\op}[1]{\hat{#1}}

\newcommand{\beq}{\begin{eqnarray}}
\newcommand{\eeq}{\end{eqnarray}}

\begin{document}
\title{Probing biological light-harvesting phenomena by optical cavities}

\author{Filippo Caruso}
\affiliation{Institut f\"{u}r Theoretische Physik, Universit\"{a}t Ulm, Albert-Einstein-Allee 11, D-89069 Ulm, Germany}

\author{Semion K. Saikin}
\affiliation{Department of Chemistry and Chemical Biology, Harvard University, 12 Oxford Street, Cambridge, MA  02138, USA}
\affiliation{Department of Physics, Kazan Federal University,18 Kremlyovskaya Street, Kazan 420008, Russian Federation}

\author{Enrique Solano}
\affiliation{Departamento de Qu\'{i}mica F\'{i}sica, Universidad del Pa\'{i}s Vasco - Euskal Herriko Unibertsitatea, E-48080 Bilbao, Spain}
\affiliation{IKERBASQUE, Basque Foundation for Science, Alameda Urquijo 36, 48011 Bilbao, Spain}

\author{Susana F. Huelga}
\affiliation{Institut f\"{u}r Theoretische Physik, Universit\"{a}t Ulm, Albert-Einstein-Allee 11, D-89069 Ulm, Germany}

\author{Al\'an Aspuru-Guzik}
\affiliation{Department of Chemistry and Chemical Biology, Harvard University, 12 Oxford Street, Cambridge, MA 02138, USA}

\author{Martin B. Plenio}
\affiliation{Institut f\"{u}r Theoretische Physik, Universit\"{a}t Ulm, Albert-Einstein-Allee 11, D-89069 Ulm, Germany}
\begin{abstract}
We propose a driven optical cavity quantum electrodynamics (QED) set up aimed at directly probing energy
transport dynamics in photosynthetic biomolecules.
We show that detailed information concerning energy transfer paths and delocalization of exciton states can be
inferred (and exciton energies estimated) from the statistical properties of the emitted photons.
This approach provides us with a novel spectroscopic
tool for the interrogation of biological systems in terms of quantum optical phenomena
which have been usually studied for atomic or solid-state systems,
e.g. trapped atoms and semiconductor quantum dots.
\end{abstract}
\maketitle
\paragraph{Introduction.--}

Following a series of fascinating non-linear optical
experiments with light-harvesting complexes (LHCs) involved in
natural photosynthesis
\cite{prok,fleming07a,Engel2010,collini,marangos}, the analysis of
exciton dynamics in biological systems has attracted a rapidly
growing interest from different scientific communities.
It has been demonstrated that the essentially lossless excitation
energy transfer (EET) in these structures
does not follow the weak-coupling F\"orster theory \cite{Foerster}
where the excitation energy incoherently hops between single molecules. Indeed,
exciton states in LHCs are frequently delocalized
over several molecules.
Several theoretical studies have analyzed EET in such pigment-protein
complexes and demonstrated that an intricate interplay between
environmental noise (dephasing) and quantum coherence is crucial
to explain the remarkably high EET efficiency (above $95\%$) from
the peripheral antenna, collecting external light, to a reaction center,
where electronic energy is transformed into chemical form
\cite{Aspuru08,Plenio08,patrick,castro08,ccdhp09,cdchp10,chp,alan1,alan2,alan3,alan4}.

On the other hand, the ability to control light-matter
interaction for single atoms~\cite{Mabuchi02} or quantum dots~\cite{Finley11} in high-quality optical cavities, e.g. photonic crystals, micropillars, and microdisk
resonators, has recently allowed the theoretical and experimental
analysis of fundamental quantum phenomena arising at a single
photon level~\cite{Khitrova2006}. In particular, quantum optical phenomena such as
the vacuum Rabi splitting, single-atom lasing, photon squeezing, anti-bunching, and
matter-light entanglement have been explored in the
strong-coupling limit, i.e. when the exchange energy rate between
the cavity mode and the trapped system is larger than the involved
decay processes~\cite{vuckovic,Murr11}. In this regime, the cavity and trapped system can no longer
be described as separate subsystems, i.e. there is a
splitting between energy eigenstates into different manifolds
(dressed states or polaritons associated to the Jaynes-Cummings model). Polariton physics has had a recent
surge in interest in the context of coupled cavity arrays
\cite{hartmann,angelakis,greentree} and in Bose-Einstein
condensation of exciton polaritons \cite{Kasprzak2006,Deng2010}. A
different, yet related, lasing process \cite{Butte2011} was also observed in organic molecular crystals in cavities
\cite{Kena-Cohen2010}.
\begin{figure}[t]
\centerline{\includegraphics[width=.4\textwidth]{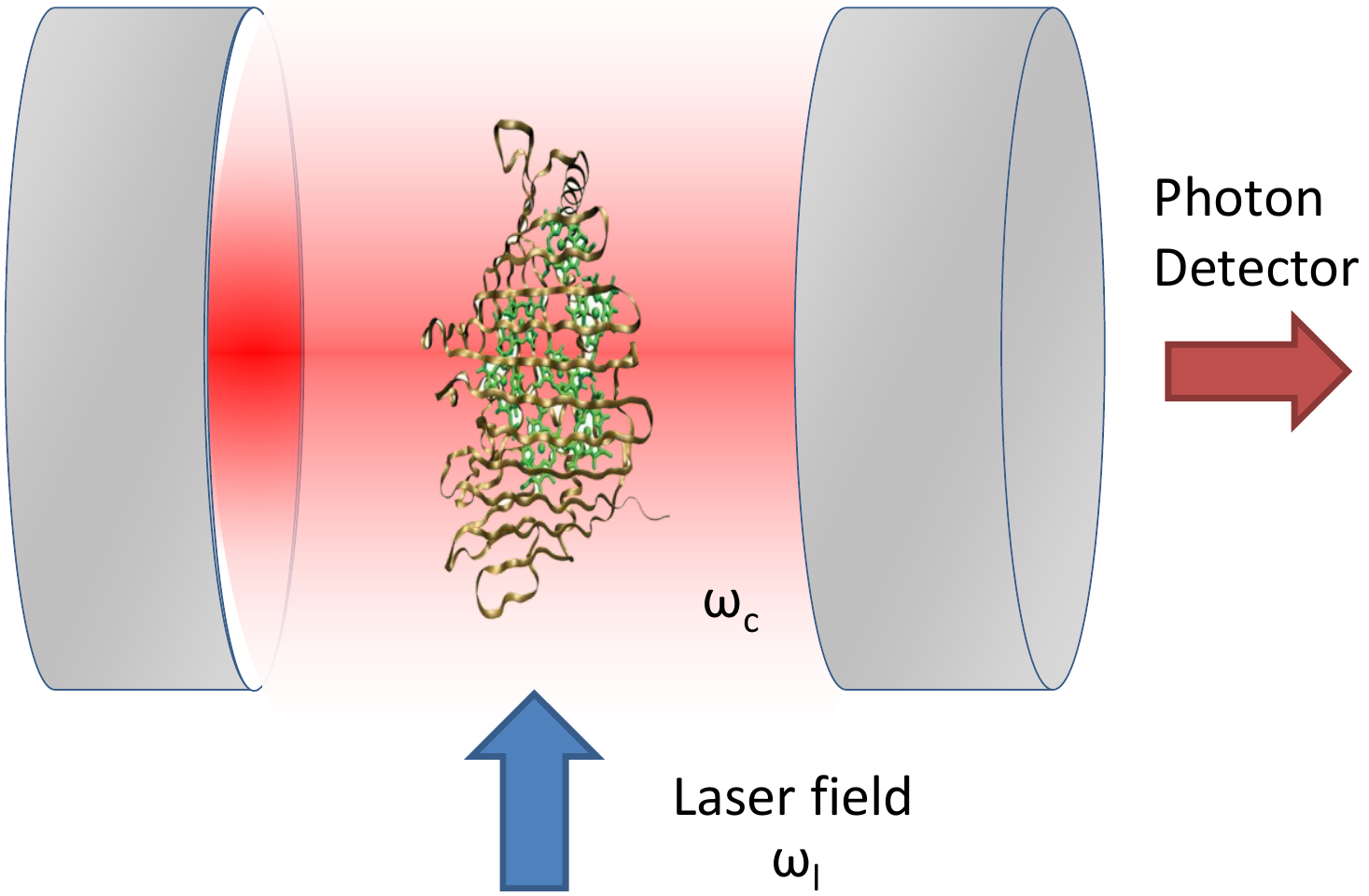}}
\caption{Artistic impression of the proposed setup: a sample of light-harvesting
complexes is confined in an optical cavity (resonance frequency $\omega_c$) with a quality factor $Q$. The sample is
subjected to an orthogonal driving laser field of frequency $\omega_l$.
Photons leaking out of the cavity are collected by a photon detector.}\label{fig1}
\end{figure}

Here, we present the extension of the rich physics
of cavity quantum electrodynamics to the investigation of the internal excitation energy transfer dynamics of bio-molecules.
In particular, we propose theoretically
an experimental scheme in which a sample of LHCs, driven
by an external laser field, is trapped inside an optical cavity.
Proof-of-principle demonstrations for
solution-phase cavity-enhanced spectroscopy were reported in \cite{mabuchi}, where
absorption measurements with Bacteriochlorophyll a (BChla) in solution
were performed.
In this Letter, the main goal is to detect quantum phenomena associated with
the emitted quantized light \cite{rempe}, in order to probe biological coherence in
photosynthetic complexes, in a different and complementary way
with respect to 2D nonlinear electronic spectroscopy used so far
in Refs. \cite{fleming07a,prok,Engel2010,collini,adolphs}.

We show that the LHC exciton structure and
energy transfer paths can be mapped onto the statistical
properties of cavity photons such as mean photon number and
second-order correlation functions and later measured through a
light mode leaking out of the cavity. Furthermore, we quantify the
amount of quantum correlations between the cavity mode and the
biological system and discuss the possibility of polariton
formation for a single LHC in a cavity.
\paragraph{Model.--}
In order to probe the exciton structure of biological systems,
we essentially propose a pump-probe scheme in which the \textit{probe}
field is substituted by the cavity mode.
The sample (\textit{system}) is assumed to be
confined inside an optical cavity with resonance frequency $\omega_c$, while energy
is injected into the system via an external laser field
(\textit{pump}) with frequency $\omega_l$. While
the pump field can be pulsed or continuous wave (CW), the probe
field (cavity mode) is always CW. We will consider the case where the pump
field is turned on at a certain time, which allows us to explore
both the stationary response to the CW field as well as the transient
excitation dynamics. Moreover, for the sake of simplicity, we consider a
collinear polarization of the laser field and the cavity mode.
The extension of the model to a pulsed excitation and cross-field
polarizations is straightforward. The LHC exciton dynamics is characterized by
monitoring the mean photon number and the second order coherence
function $g^{(2)}(0)$ of the light leaking out of the cavity
(probe).
The coherent interaction between the cavity mode and the LHC chromophores can
be written as
\begin{equation}
H_c =  \sum_i g \mu_i \left( \sigma_i^{+}
a + \sigma_i^{-}  a^\dagger \right), \label{H_FMO_cavity}
\end{equation}
where $g$ is the coupling constant, $\mu_i$ is the projection
of the transition dipole moment of a single chromophore to the
polarization of the cavity mode, $\op{\sigma}_i^{\pm}$ are rising
and lower operators for chromophore excitations, and
$\op{a}^\dagger$, $\op{a}$ denote the cavity photon creation/annihilation
operators. See the supplementary information (SI) for additional details concerning the model.
The rotating wave approximation utilized in
Eq.~(\ref{H_FMO_cavity}) is valid for the coupling regime we
consider here. For the sake of simplicity we analyze photon
statistics in a single polarization mode in the cavity. In
general, there are two degenerate polarization modes. However, the
interaction between them is mediated by the LHC and the associated
effects in the photon statistics are of higher order in the
LHC-cavity coupling as compared to the single photon emission by
the LHC. The coupling coefficient $g$ is given by $g=
\sqrt{\frac{\omega_c}{2\varepsilon V}}$, where $\hbar=1$, $V$ is
the mode volume, and $\varepsilon$ is the permittivity of the
medium between the cavity mirrors. In the following, we assume
$\varepsilon = 2$, which is approximately the
dielectric constant of a solvent at optical frequencies.
The cavity properties can be described by the quality factor $Q$
and the mode volume $V$.  While the former parameter determines
the cavity photon escape rate as $ \kappa = \omega_c/Q$,
the latter one controls the coupling with the LHC. Experimentally
achievable values of the quality factor can be as large as $Q = 10^5-10^7$
\cite{Khitrova2006,CNRS_PRL2010,FP_cavity_NJP2010}. The fundamental limit
of the mode volume is determined by the photon wavelength $V_f = (\lambda/2)^3$
and is dependent on the dielectric constant of the medium. While the mode
volume in photonic crystal cavities is about $\lambda^3$
\cite{Khitrova2006}, to the best of our knowledge the state-of-the-art mode volume,
demonstrated for a Fabry-Perot cavity design \cite{FP_cavity_NJP2010},
which can be used for molecular spectroscopy, is $V_{\rm exp} \approx 5.5 \lambda^3$.
This value corresponds to $g \approx 0.015$~cm$^{-1}$/D coupling between the LHC
exciton and the cavity vacuum state for $\lambda_{\rm vacuum} = 800$~nm. A more optimistic estimation
for the coupling constant obtained for the fundamental limit of
the mode volume, i.e. $V_f \approx 0.023$~$\mu$m$^3$, is $g_{\rm th}  \approx
0.1$~cm$^{-1}/D$. In the following, we analyze the results corresponding to the
latter case in order to enhance the fundamental issues of molecular
spectroscopy in a cavity we want to stress, while
we also discuss (in SI) the state-of-the-art case to answer the question
of which effects can be observed at the present stage of technological advances.
As an example of an LHC, we choose the Fenna-Matthews-Olson (FMO)
pigment-protein complex involved in the natural photosynthesis of green-sulphur bacteria.
The FMO subunit has seven~\cite{note} strongly coupled
bacteriochlorophyll (BChl) molecules (sites) \cite{prok,fleming07a}. The
Hamiltonian of the complex can be written as
\begin{equation}
H_s = \sum_j \omega_j \sigma_j^{+} \sigma_j^{-}
        + \sum_{j\neq l} v_{j,l} (\sigma_j^{-}
        \sigma_{l}^{+} + \sigma_j^{+} \sigma_{l}^{-}) \; ,
        \label{FMO_H}
\end{equation}
where only lowest electronic transitions of the molecules are
taken into account, $\omega_j$ is the site
energy, and $v_{j,l}$ denotes the coherent coupling between the
corresponding sites. The single exciton transitions in the FMO
complex are estimated to be in the range of $12150-12750$~cm$^{-1}$
\cite{Fleming2005,adolphs}.
The excitation dynamics in the FMO-cavity system can be described by a
general Liouville-von Neumann equation for the density matrix
\begin{equation}
\dot{\rho} = -i [H,\rho] + \mathcal{L}(\rho),
\end{equation}
where the total Hamiltonian $H$ includes the FMO Hamiltonian
Eq.~(\ref{FMO_H}), the cavity Hamiltonian $H_{\rm ph} = \omega_c a^{\dagger}a$,
the interactions between FMO and the cavity Eq.~(\ref{H_FMO_cavity}),
and also the coupling between the FMO and an external field $E(\omega_l,t)$ given by
 \begin{equation}
 H_{l}(t) = - \sum_i \mu_i E(\omega_l,t) \sigma^{+}_i + h.c.
 \label{laser_H}
 \end{equation}
The dephasing and relaxation channels in the isolated FMO system
impose an irreversible dynamics of an initially created exciton
state. The simplest way to introduce these noise processes into
the dynamics is using a Lindblad form for the term $\mathcal{L}(\rho)$ where
the environmental noise is described by the site energy
fluctuations (phenomenological Haken-Strobl noise) \cite{haken}
with dephasing noise strength $\gamma$. This analysis will bring out the basic principles of our proposal. Extensions to non-Markovian effects which can arise from strong coupling and/or the form of the environment spectral function \cite{ccdhp09,cdchp10,aki,patrickNJP,prior2010,thorwart2009}
as well as other cavity setups involving additional laser fields, the role of multiple excitations, and other generalizations
will be investigated in a forthcoming paper.
\paragraph{Results and discussion.--}
\begin{figure}[t]
\centerline{\includegraphics[width=.5\textwidth]{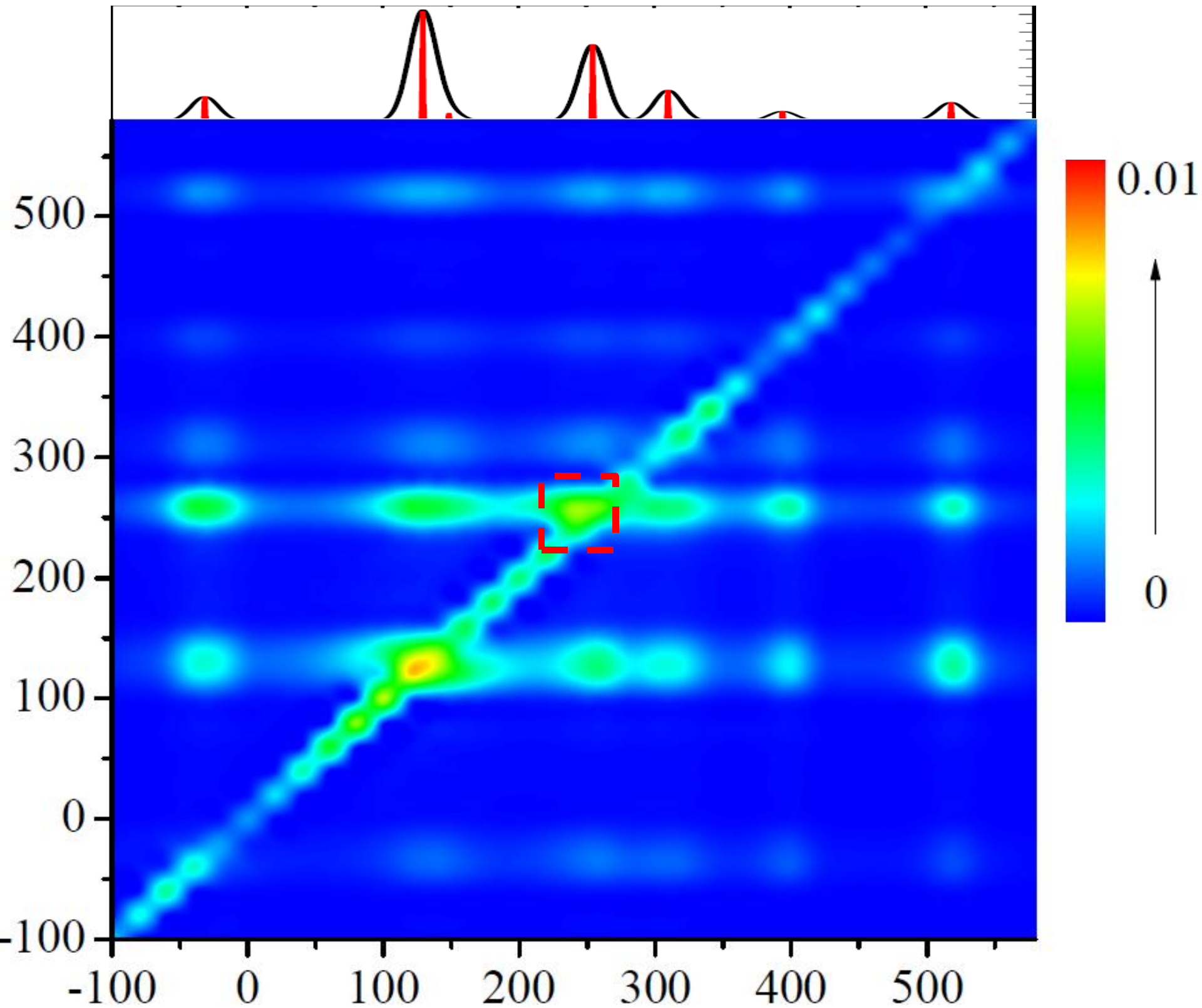}} \caption{
Stationary mean photon number of the cavity mode as a function of
the cavity ($Q=10^4$) resonance frequency $\omega_c$ (vertical
axis) and the pump laser frequency $\omega_l$ (horizontal), both
in units of cm$^{-1}$, but shifted by 12195 cm$^{-1} \sim
820~\mathrm{nm}$, in the case of dephasing rate $\gamma=10 \
cm^{-1}$. The diagonal peaks correspond to the FMO exciton
energies, while the off-diagonal peaks are related to energy
transfer between different excitons. The asymmetry of the 2D spectrum, with respect to the main diagonal ($\omega_l = \omega_c$),
is due to the different strengths of the laser and the cavity fields.
Top Inset: computed spectrum of electronic excitations the FMO
complex. The peaks corresponding to the FMO exciton energies are
broadened by the $\gamma=10$~ cm$^{-1}$ linewidth. The red box is analyzed in Fig. \ref{fig3}.}\label{fig2}
\end{figure}
\begin{figure}[t]
\centerline{\includegraphics[width=.5\textwidth]{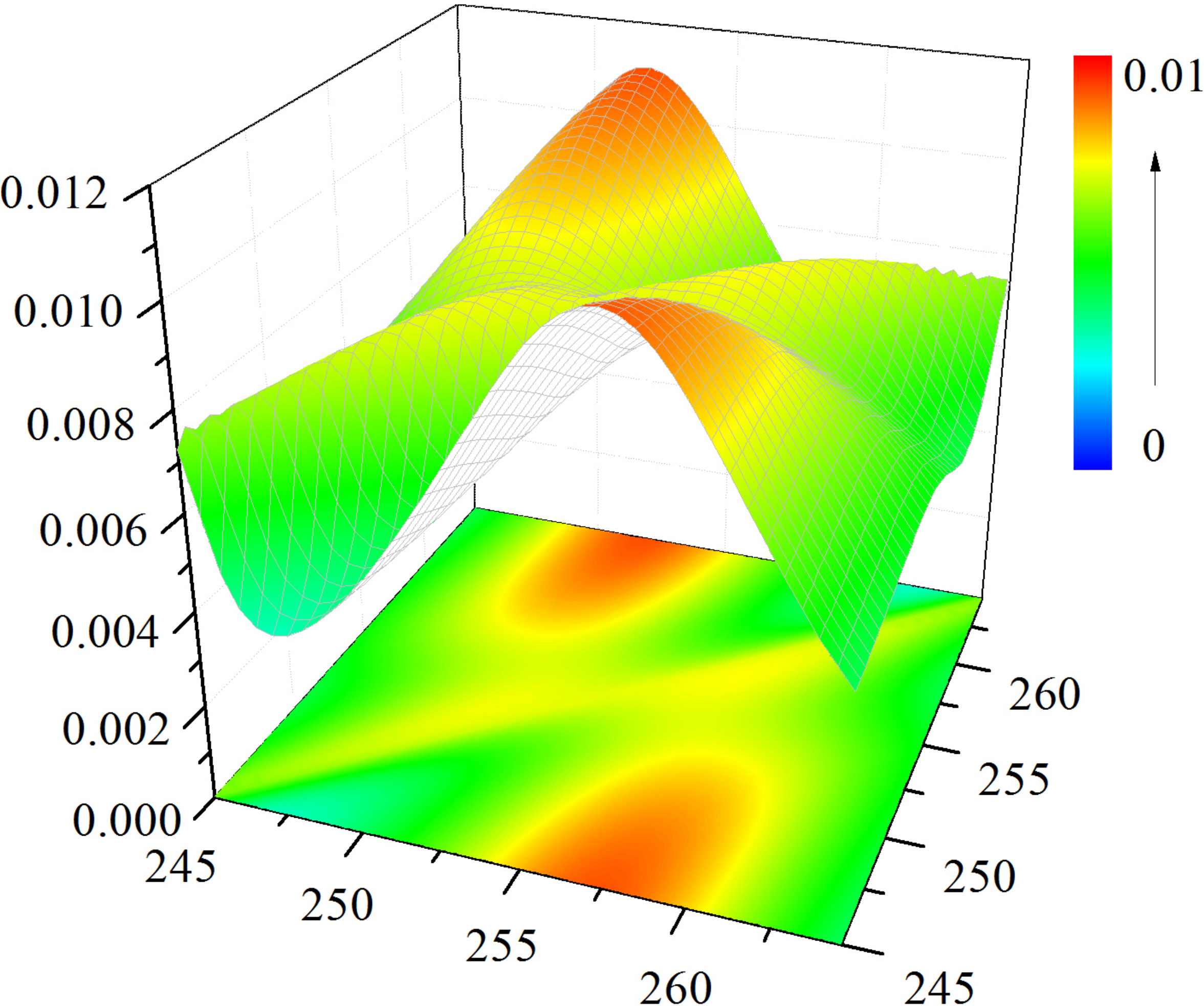}}
\caption{Mean photon number of the stationary cavity mode as a
function of the cavity ($Q=10^4$) resonance frequency $\omega_c$ (bottom axis)
and the pump laser frequency $\omega_l$ (right axis), both in units of
$cm^{-1}$ -- but shifted by $12195 \ cm^{-1} \sim 820~\mathrm{nm}$
-- in the finely tuned range $[245,\ 265]\ cm^{-1}$ (red box in Fig. \ref{fig2}). One can clearly observe an
anti-crossing effect. }\label{fig3}
\end{figure}
In Fig.~\ref{fig2}, we show the mean photon number in the cavity,
in the stationary state, as a function of the driving field and
the cavity mode frequencies together with the electron excitation
spectrum of FMO. This 2D map is obtained 45~ps after a strong
laser driving field of intensity $I \approx 110$~KW/cm$^2$ has
been turned on. The estimated values of electronic transition dipoles
of the FMO chromophores are in the range of $3-14$~D (see SI for the details of the
model). Thus, the maximal coupling energy between the driving
field and the excitons is of the order of $1-7$~cm$^{-1} =
30-210$~GHz and the exciton-cavity coupling is about $5$ times
smaller. The cavity photon population saturates to a stationary
regime on $20$~ps timescale showing a set of peaks that are clearly
associated with the exciton frequencies of the FMO complex.
Indeed, in analogy to 2D spectroscopy
\cite{fleming07a,prok,Engel2010,collini,adolphs}, diagonal peaks
are in correspondence of the FMO exciton energies, and
off-diagonal features appear due to energy transfer between
different exciton states.
The diagonal line in the spectra is due to both a coherent
and an incoherent transfer of the photons from the laser field
into the cavity. The former process is similar to the Raman
scattering and goes through coherence pathways without generating
an exciton population in LHC, while the latter process can be
considered as a light absorption by the LHC with the
following spontaneous emission of a same-frequency photon into the cavity.
The horizontal lines in the
spectra (off-diagonal peaks) are due to population
transfer between different excitons. Weaker optical fields would result in a similar set of
peaks, but the stationary regime would be achieved on a longer
timescale. For the particular case shown in Fig.~\ref{fig2}, the
noise strength is $\gamma=10$~cm$^{-1} \sim 2$~ps$^{-1}$
\cite{Engel2010,hayes2010}. Qualitatively, this picture remains
valid for the values of $\gamma$ within the range $1-100~cm^{-1}$
- see SI for more data. In all our simulations, we include an
isotropic averaging of the measured quantities over completely
random orientations of FMOs in the cavity. In presence of an
oriented sample these quantum features would be further enhanced.
\begin{figure}[b]
\centerline{\includegraphics[width=.47\textwidth]{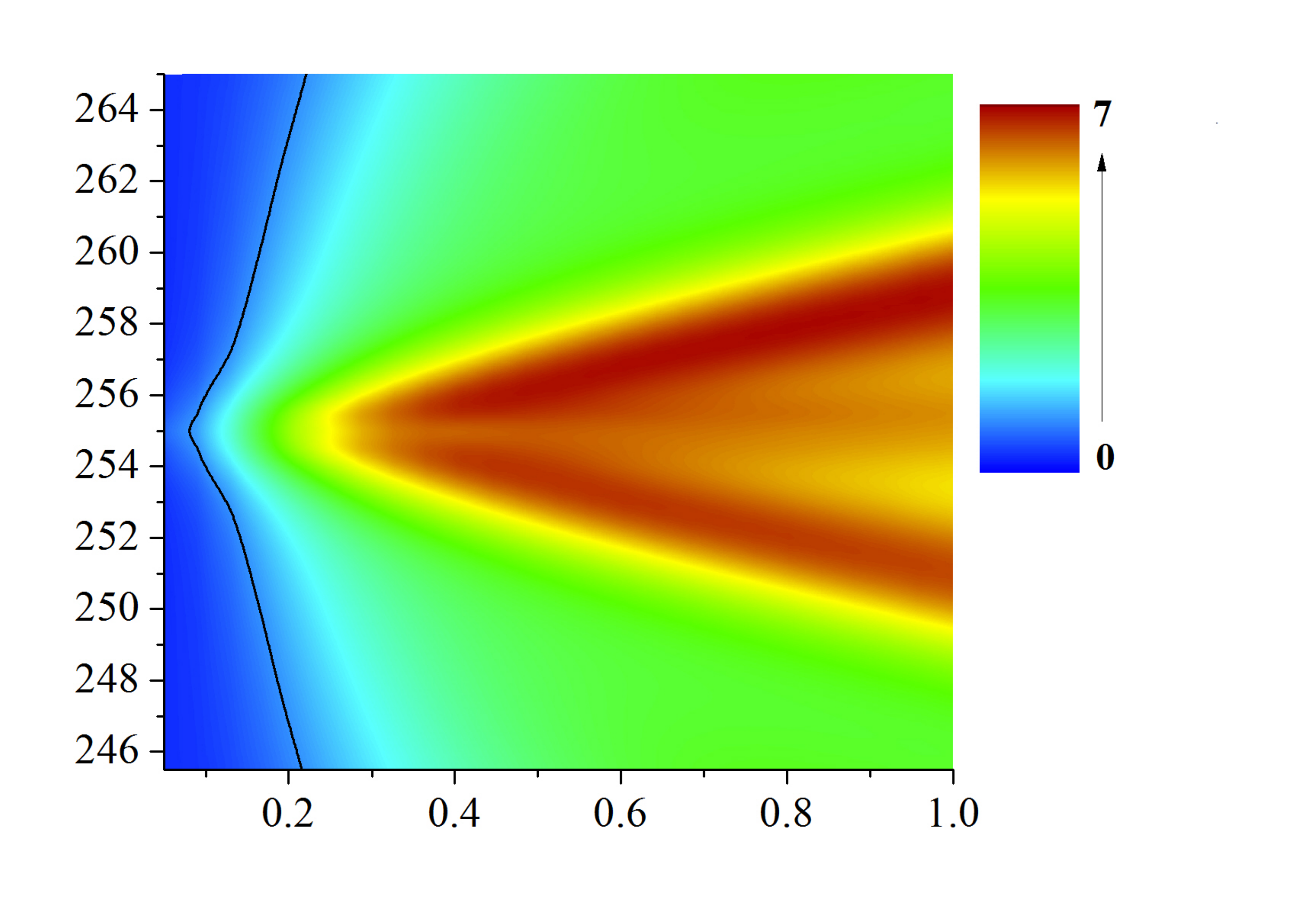}}
\caption{Second-order coherence function at time delay zero $g^{(2)}(0)$ of the stationary cavity mode as a
function of the laser field coupling (horizontal axis) and the cavity ($Q=10^4$) resonance frequency $\omega_c$ (vertical)
- in units $cm^{-1}$ - for pump laser frequency $\omega_l=-\omega_c + 510 cm^{-1}$.
The black line corresponds to $g^{(2)}(0)=1$.}\label{fig8}
\end{figure}
Stronger, albeit (biological)-structure preserving, laser driving fields, can
result in a fine structure of a diagonal peak. This effect is
similar to a level anti-crossing or a hole
burning -- see Fig.~\ref{fig3}. Indeed, the relatively strong
coupling with the pump laser field leads to a splitting of the
exciton level into two dressed states corresponding to two new
peaks in the mean photon number 2D map.
\begin{figure}[t]
\centerline{\includegraphics[width=.47\textwidth]{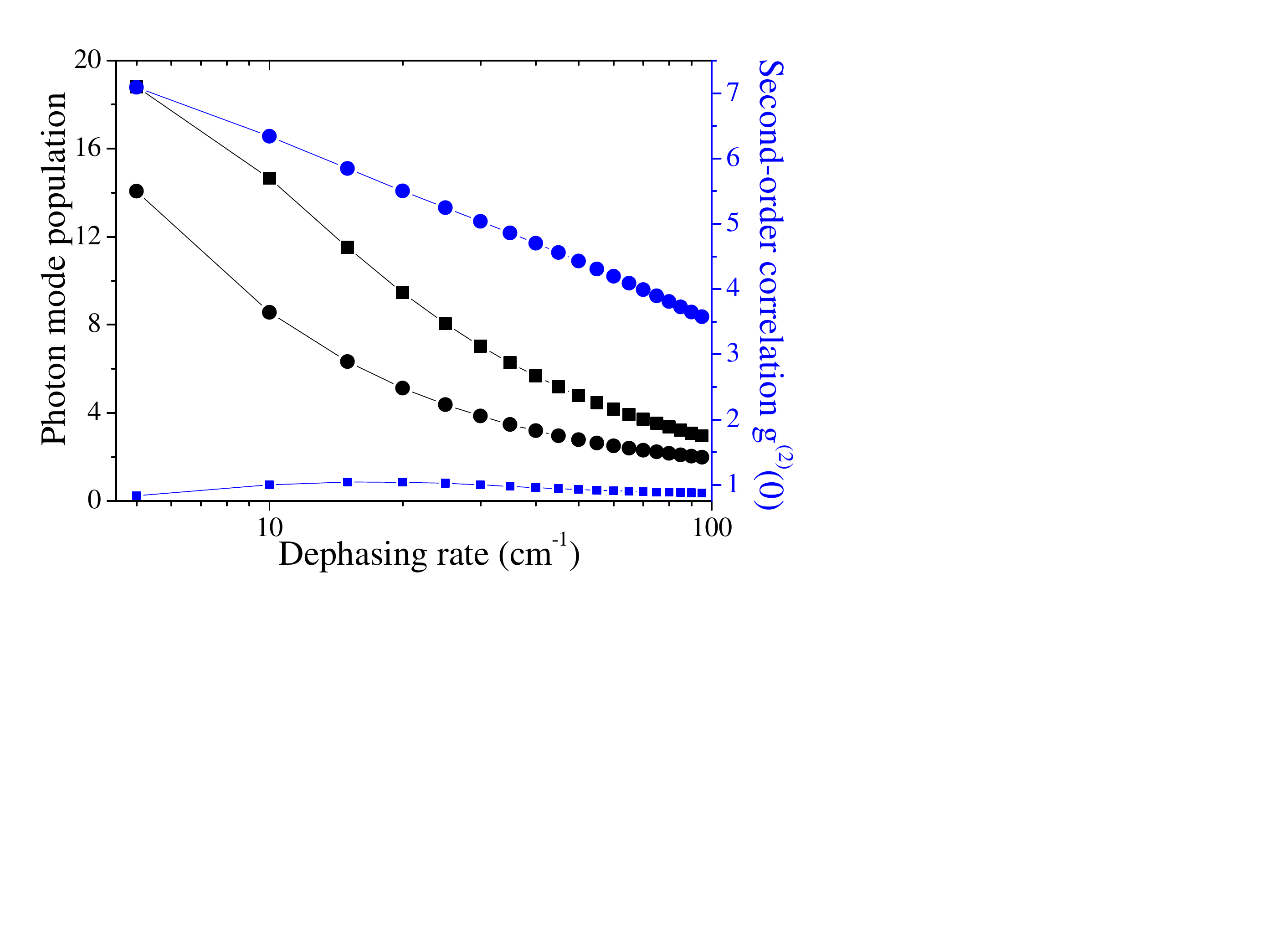}}
\caption{Photon mode population (rescaled by a factor $10^{-3}$) and second-order coherence function at zero time delay $g^{(2)}(0)$ of the stationary cavity mode as a function of the dephasing rate ($cm^{-1}$), for a diagonal (circles) and off-diagonal (square) peak in the mean photon number 2D map in Fig. \ref{fig2}.}\label{fig20}
\end{figure}
In order to characterize quantum properties of the generated
cavity photon state in its stationary state, we compute the
second-order photon coherence function~\cite{MandelWolf} at zero
time delay $g^{(2)}(0) = \frac{\langle a^\dagger a^\dagger
aa\rangle}{\langle a^\dagger a \rangle^2}$ -- see Fig. \ref{fig8}.
In general, we obtain a non-classical photon
state generation ($g^{(2)}(0) < 1$) for most of the laser and cavity
frequencies except of the diagonal line with $\omega_c = \omega_l$
where a coherent state ($g^{(2)}(0) = 1$) and a
thermal state ($g^{(2)}(0) > 1$) can be observed.
More data on $g^{(2)}(0)$ in different regimes can be found in SI. Moreover, it turns out, as in Fig. \ref{fig20}, that this quantity $g^{(2)}(0)$ depends on the amount of dephasing noise in the dynamics and this might become a tool to estimate the strength of interaction between the biological molecule and the external environment. Notice also that the photon mode population is more sensitive to the dephasing noise, as compared to $g^{(2)}(0)$. In addition, the new scheme we are proposing allows us also to generated non-classical states of light by means of the coherent exciton dynamics.
Finally, as possible signature of polariton formation,
we quantify the amount of quantum correlations between
the FMO sample and the cavity mode, as measured by the logarithmic
negativity \cite{plenio05}. It turns out that this quantity
reaches a maximum at about $800 \ \mathrm{fs}$ and around the main
two diagonal peaks in the 2D map in Fig. \ref{fig2} -- see Fig. \ref{fig4}. Note that the lifetime of these correlations is around $2$~ps$^{-1}$, which is the time scale of the fastest decay process due to the presence of dephasing noise. Therefore, the coupling between the quantum cavity and the confined sample leads to the
creation of non-classical correlations between these two systems.
A different scheme towards entangling two FMO samples
(entangled polaritons) in separate cavities is also discussed in
SI.
\begin{figure}[h]
\centerline{\includegraphics[width=.46\textwidth]{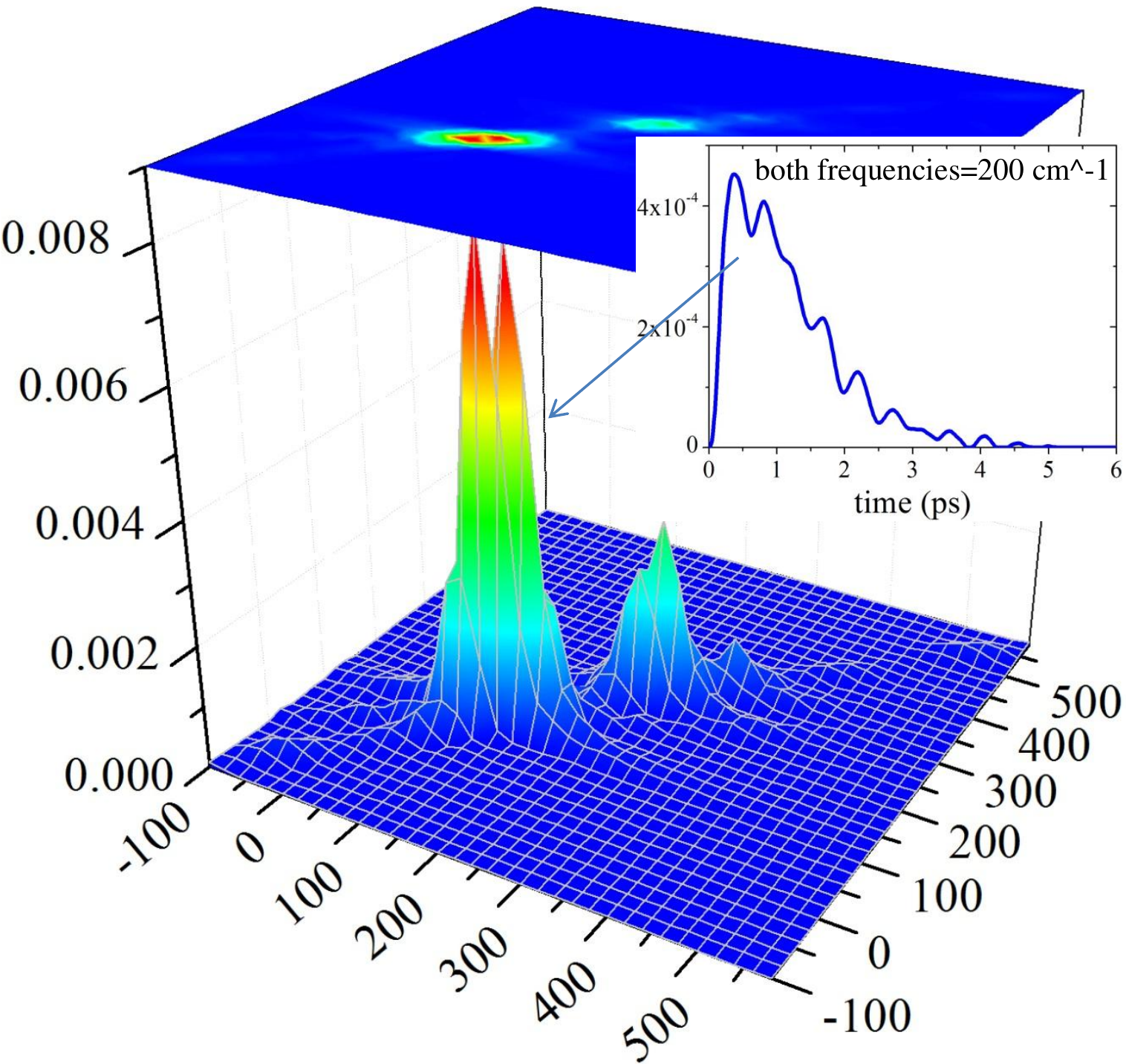}}
\caption{Measure of quantum correlation (log-negativity) between
the cavity mode and the FMO sample after $\sim 800 \mathrm{fs}$, as a function
of the cavity ($Q=10^4$) resonance frequency $\omega_c$ and the pump laser frequency $\omega_l$,
both in units of cm$^{-1}$, but shifted by 12195 cm$^{-1}$, in the case of dephasing rate $\gamma=10 \ cm^{-1}$.
Inset: Time evolution of the log-negativity measure for
$\omega_c=\omega_l=200 cm^{-1}$.}\label{fig4}
\end{figure}
\
\newline
\paragraph{Conclusions and outlook.--}
We have extended the rich physics of cavity quantum electrodynamics to biological molecules, in particular light-harvesting complexes involved in natural photosynthesis. As an example, we have considered a sample of the Fenna-Matthews-
Olson (FMO) pigment-protein complex inside an optical cavity and driven transversally by a laser field. Our main result is that the emission spectrum from the coupled FMO-cavity system reflects coherent energy transfer into the cavity through delocalized
exciton states, as well as exciton dynamics within the FMO complex. Moreover, we have found that a strong laser field driving the
exciton dynamics in the protein complex can burn a hole in the emission spectrum resulting in an additional
structure of resonance peaks. Finally, the generation of quantum states of light inside the cavity, due to the interaction with the confined biological complexes, was also observed.
We believe that this novel spectroscopic tool can efficiently probe both static and dynamical (system-environment) properties of natural and artificial photosynthetic structures and
provide additional experimental support for the coherent dynamics unveiled by non-linear spectroscopy experiments. This combined evidence is expected to shed further light towards the elucidation of the role that quantum coherence may play in the remarkably robust energy transport phenomena involved in natural photosynthesis.

This work was supported by DARPA-QUBE, DTRA Contract No HDTRA1-10-1-0046, the EU projects Q-ESSENCE
AQUTE, CORNER, HIP, SOLID, and CCQED, the Spanish Ministry of Education (project MICINN FIS2009-12773-C02-01), the Basque government (project IT472-10), the SFB/TRR21, the EU Marie-Curie Programme, and the
Alexander von Humboldt Foundation.

\
\newpage

\begin{widetext}
\begin{center}
\Large{\textbf{Supplementary Information}}
\end{center}
\end{widetext}
\section{The Model}
In the site basis, the Hamiltonian of the FMO pigment-protein complex, $H_s $ (in the main text), is given by
\begin{equation}
        \left(\!\!\begin{array}{rrrrrrr}
         215   & \!-119.0 & 6.8  & -7.2  &   8.6 & -18.1 &  -14.9 \\
        \!-119.0 &  305 & 36.9 & 9.3   &   2.1 &   16.3 &   6.9 \\
           6.8 &   36.9 &  0 & -69.8 &   -1.5 &  -11.5 &   3.7 \\
          -7.2 &    9.3 &\!-69.8 & 200 &\! -78.3 &\! -21.3 &  -75.9\\
           8.6 &    2.1 &  -1.5 & \!-78.3 & 425 &  110.3 &  -6.4 \\
         -18.1 &    16.3 & -11.5 & -21.3 &  110.3 & 315 &  43.0 \\
          -14.9 &    6.9 &  3.7 & -75.9 &  -6.4 &  43.0 & 265
          \end{array}\!\!
        \right)
        \label{hami} \nonumber
\end{equation}
where the diagonal elements are the site energies -- shifted from
the base line $\omega_{\rm base} = 12195$~cm$^{-1}$ corresponding
to a wavelength of $\cong 820~\mathrm{nm}$ -- while the
off-diagonal elements are the inter-site coupling rates (all
numbers are given in units of $\text{cm}^{-1}=1.988865\cdot
10^{-23}~\mathrm{Nm} = 1.2414 \ 10^{-4}~\mathrm{eV}$).
The off-diagonal terms of the Hamiltonian were calculated in a dipole-dipole approximation and the site energies were taken from the Poisson-Boltzmann quantum chemistry model of Ref.~\cite{Renger2008}. The intensities of the electronic transitions in the computed absorption spectrum (top inset in Fig.~1 of the main text) are proportional to $|\mu_z|^2$, and the spectrum is averaged over
different spatial orientations of the FMO complex.
The dissipation and dephasing caused by the surrounding environment can be described by the following local Lindblad terms:
 \be
\mathcal{L}^S_{diss}(\op{\rho}) = \sum_{j=1}^{7} \Gamma_j/2
[-\{\op{\sigma}_j^{+} \op{\sigma}_j^{-},\op{\rho}\} + 2
\op{\sigma}_j^{-} \op{\rho} \op{\sigma}_j^{+} ] \; , \ee \be
\mathcal{L}^S_{deph}(\op{\rho}) = \sum_{j=1}^{7} \gamma_j/2
[-\{\op{\sigma}_j^{+} \op{\sigma}_j^{-},\op{\rho}\} + 2
\op{\sigma}_j^{+} \op{\sigma}_j^{-} \op{\rho} \op{\sigma}_j^{+}
\op{\sigma}_j^{-}] \; , \label{locdeph}        \ee with $\Gamma_j$
and $\gamma_j$ being the dissipative and dephasing rates at the
site $j$, respectively, $\op{\sigma}_j^{+}$ ($\op{\sigma}_j^{-}$)
being the raising (lowering) operators for site $j$.
In the following, we choose for simplicity uniform dephasing
rates, i.e. equal $\gamma_j$ and so simply labeled as $\gamma$.
Moreover, here we neglect the presence of dissipation inside the
FMO complex, because the $1~\mathrm{ns}$ excitation
lifetime \cite{adolphs06} is much longer than the time scale we look at. Let us
point out that some dissipation is also present indirectly through
the interaction with the cavity which has a damping channel,
described by the Lindbladian term:
\begin{equation}
\mathcal{L}^C_{diss}(\op{\rho}) = \Gamma_C/2 \left( - \left\{
\op{a}^{\dag}\op{a}, \op{\rho} \right\} +2 \op{a}
\op{\rho} \op{a}^{\dag} \right) \; ,
\end{equation}
where $\Gamma_C$ being the damping rate of the cavity modes due to some dissipative processes, i.e. $\Gamma_C=\frac{\omega_C}{Q}$ with $Q$ being
the quality factor of the cavity.
For a self-consistent model we use the same transition dipoles to
construct interactions between BChl molecules in the FMO
Hamiltonian and also for the coupling with the cavity and external
fields. The positions and the relative directions of the dipoles
are extracted from the structure of FMO complex \cite{PDB}. The
positions $R_i$ of the $7$ sites (BChl), taken as a middle point between 4
Nitrogen atoms in each BChl, are
\begin{eqnarray}
[R_i]=\left( {\begin{array}{*{20}c}
      53.08 &  58.26 & 20.64\\
      56.04 & 54.79 & 32.40\\
      49.57 & 44.77 & 45.39\\
      38.81 & 42.25 & 43.06\\
      34.15 & 47.78 & 31.26\\
      41.44 & 47.82 & 22.61\\
      47.53 & 43.95 & 33.22
      \end{array}}\right).
\end{eqnarray}

The directions the three components (in some reference frame) of the $7$ induced-transition dipoles $\mu_i$ are
\begin{eqnarray}
\frac{[\mu_i]}{\mu_0}=\left( {\begin{array}{*{20}c} -0.026 & 0.286 & -0.958 \\
           -0.752 & 0.601 & -0.271\\
           -0.935 & 0.061 &  0.349\\
           -0.001 & 0.393 & -0.919\\
           -0.739 & 0.672 &  0.048\\
            0.859 & 0.371 & -0.353\\
            0.176 &-0.042 & -0.983 \end{array}}
            \right),
\end{eqnarray}
where the absolute values of the transition dipoles are taken as a
phenomenological parameter $\mu_0=6$~D, which accounts for effects
of screening and induced charges \cite{Renger2008}. The exciton energies corresponding to the Hamiltonian (\ref{hami}) are $E_{1-7} = [-31.0, 129.4, 148.7, 254.6, 310.1, 394.5, 518.6]$~cm$^{-1}$ and the absolute values of the electronic transition dipoles of the FMO chromophores are $|\mu_{1-7}| = [6.4, 13.8, 3.7, 11.3, 7.2, 4.0, 5.6]$~D.
\begin{figure}[t]
\centerline{\includegraphics[width=.435\textwidth]{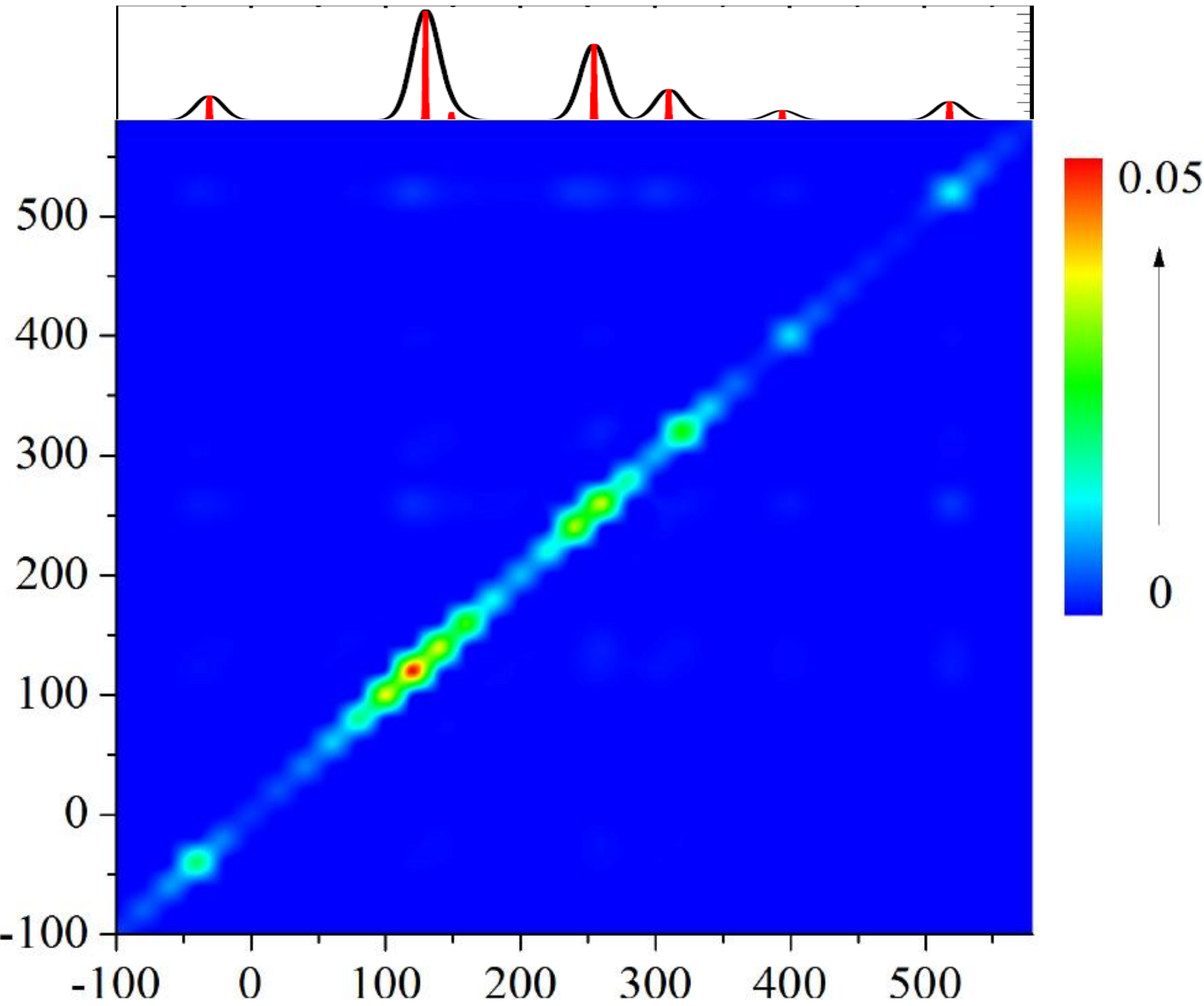}}
\centerline{\includegraphics[width=.45\textwidth]{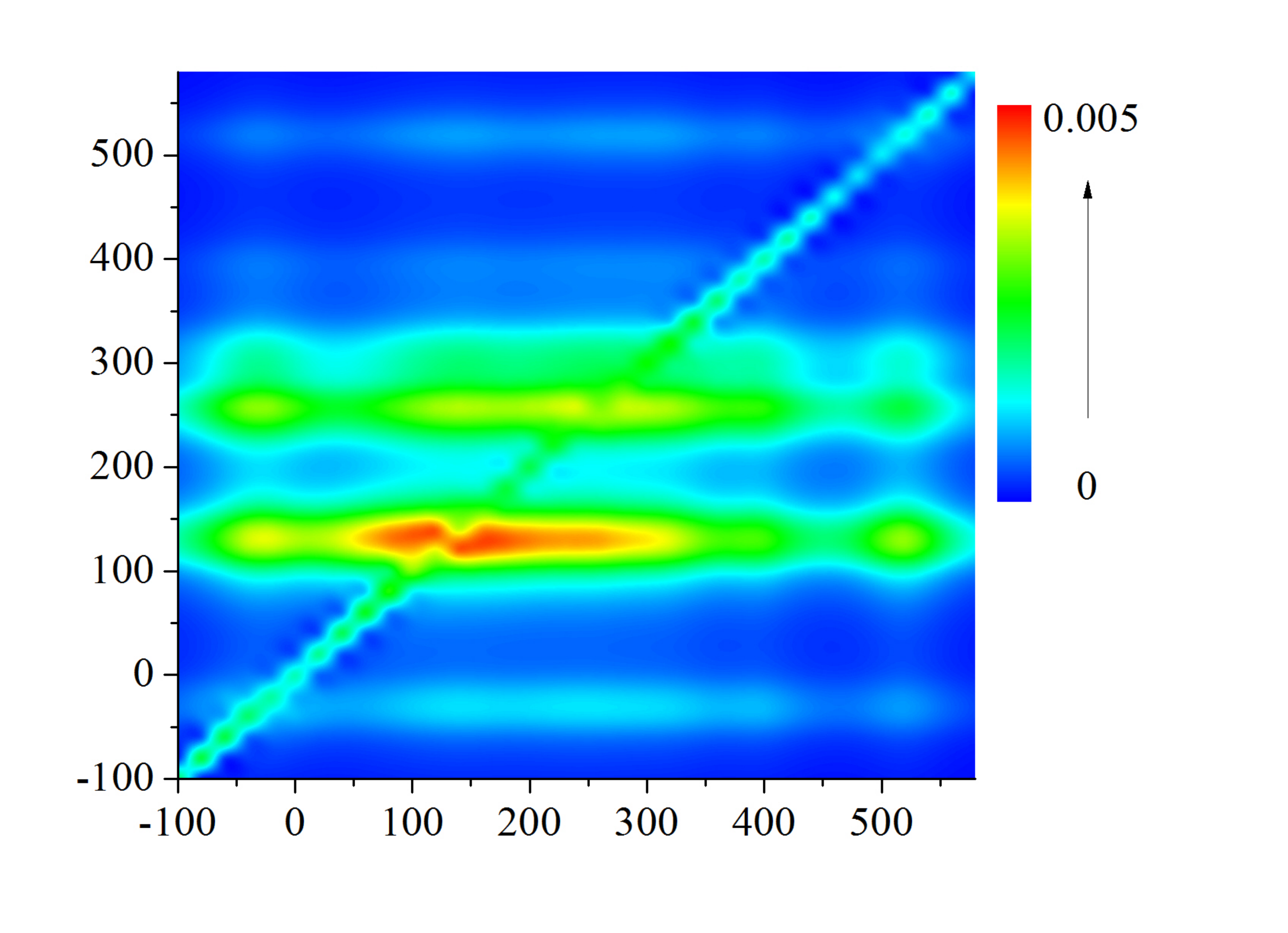}}
\centerline{\includegraphics[width=.45\textwidth]{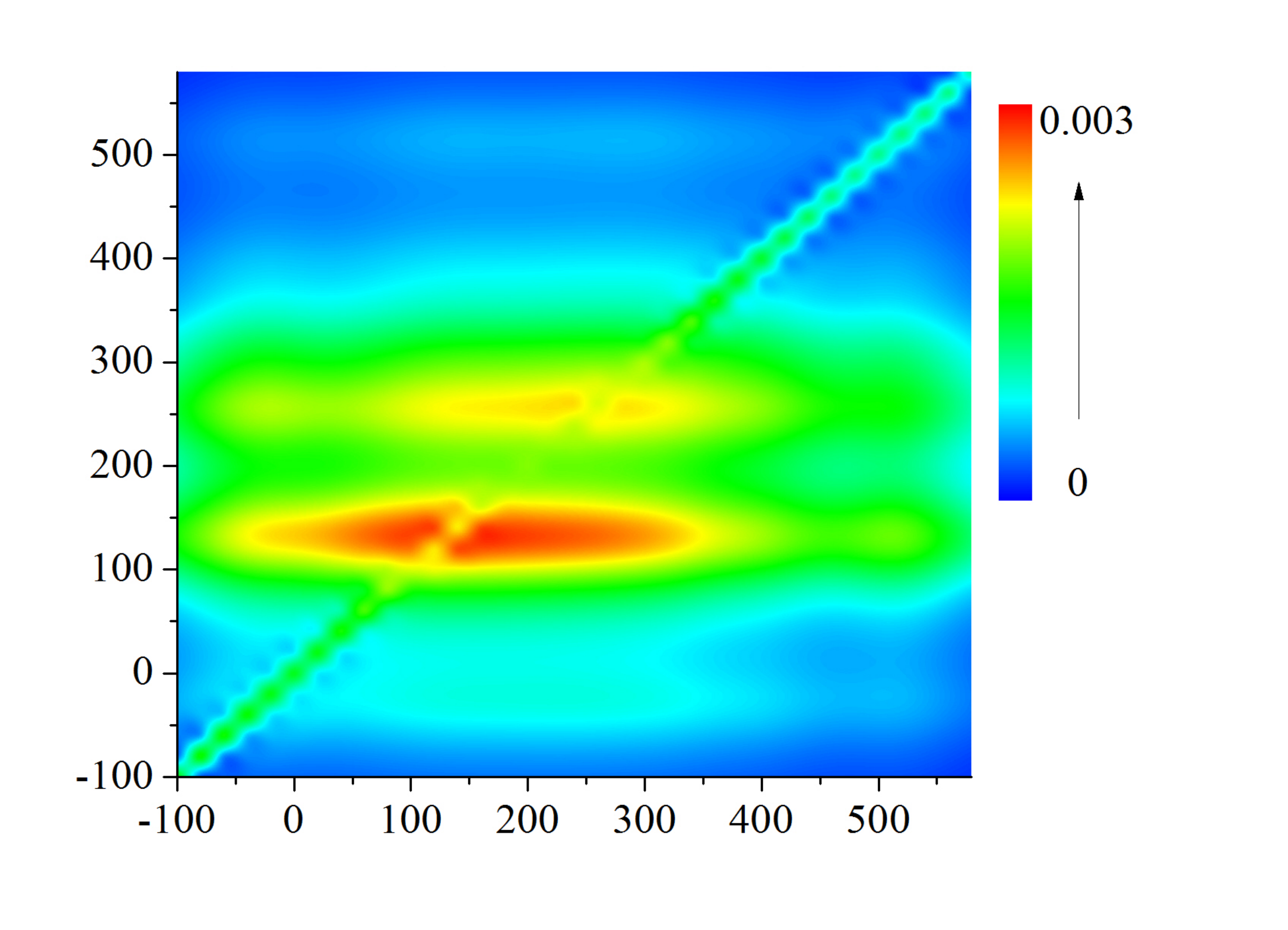}}
\caption{
Stationary mean photon number of the cavity mode as a function
of the cavity ($Q=10^4$) resonance frequency $\omega_c$ (vertical
axis) and the pump laser frequency $\omega_l$ (horizontal),
both in units of cm$^{-1}$, but shifted by 12195 cm$^{-1} \sim
820~\mathrm{nm}$, for dephasing rates $\gamma=1,\ 50,\ 100 \ cm^{-1}$ (from top to bottom).}\label{fig13}
\end{figure}

Finally, we assume a continues-wave laser pumping exciton transitions in
the LHC. The laser field is coupled to the cavity modes
through the trapped bio-sample only. The value of the external (pump) field can be estimated as
\begin{equation}
E = \sqrt{\frac{2I}{c\varepsilon_0}},
\end{equation}
where $I$ is the laser field intensity, $c$ is the speed of light,
and $\varepsilon_0$ is the vacuum permittivity. For example, the
intensity of a CW laser field used in Raman spectroscopy is about
5~mW. In these experiments the laser beam is focused on a spot of
a size of 5~$\mu$m$^2$. This corresponds to $I=100$~kW/cm$^2$ or
the field $E=27\cdot10^5$~V/m~$=0.45$~cm$^{-1}/D$. This value is
much smaller than the peak field value used in femtosecond pulse
lasers. However, a CW would result in a heat accumulation, thus
stronger fields could damage a sample.
\begin{figure}[t]
\centerline{\includegraphics[width=.47\textwidth]{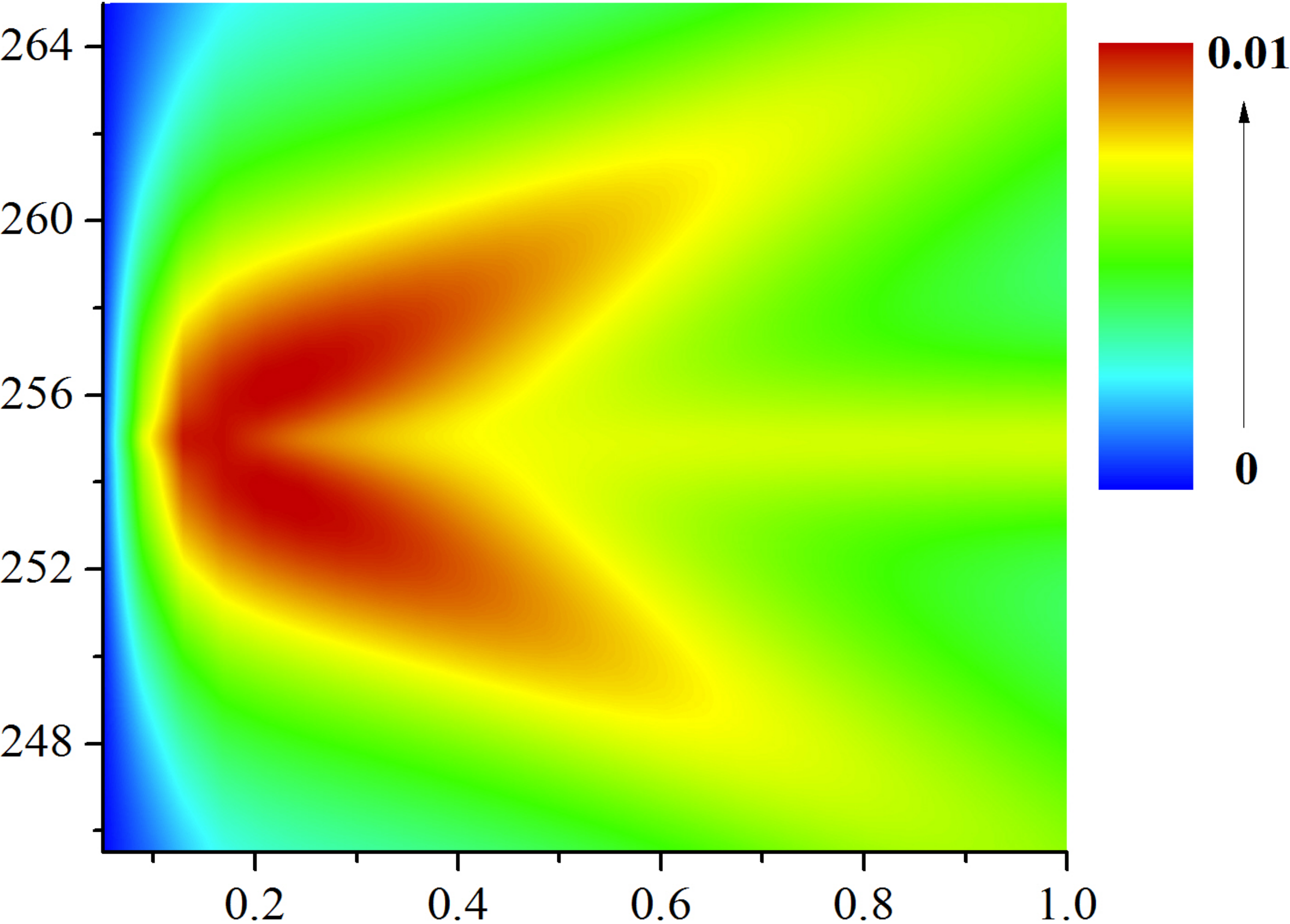}}
\caption{Average photon number of the stationary cavity mode as a
function of the laser field coupling (horizontal axis) and the cavity ($Q=10^4$)
resonance frequency $\omega_c$ (vertical) - in units $cm^{-1}$ -
for pump laser frequency $\omega_l=-\omega_c + 510 cm^{-1}$,
which corresponds to an cross-diagonal cut of a diagonal peak (red box in Fig. 2 of the main manuscript).}\label{fig7}
\end{figure}
\begin{figure}[t]
\centerline{\includegraphics[width=.48\textwidth]{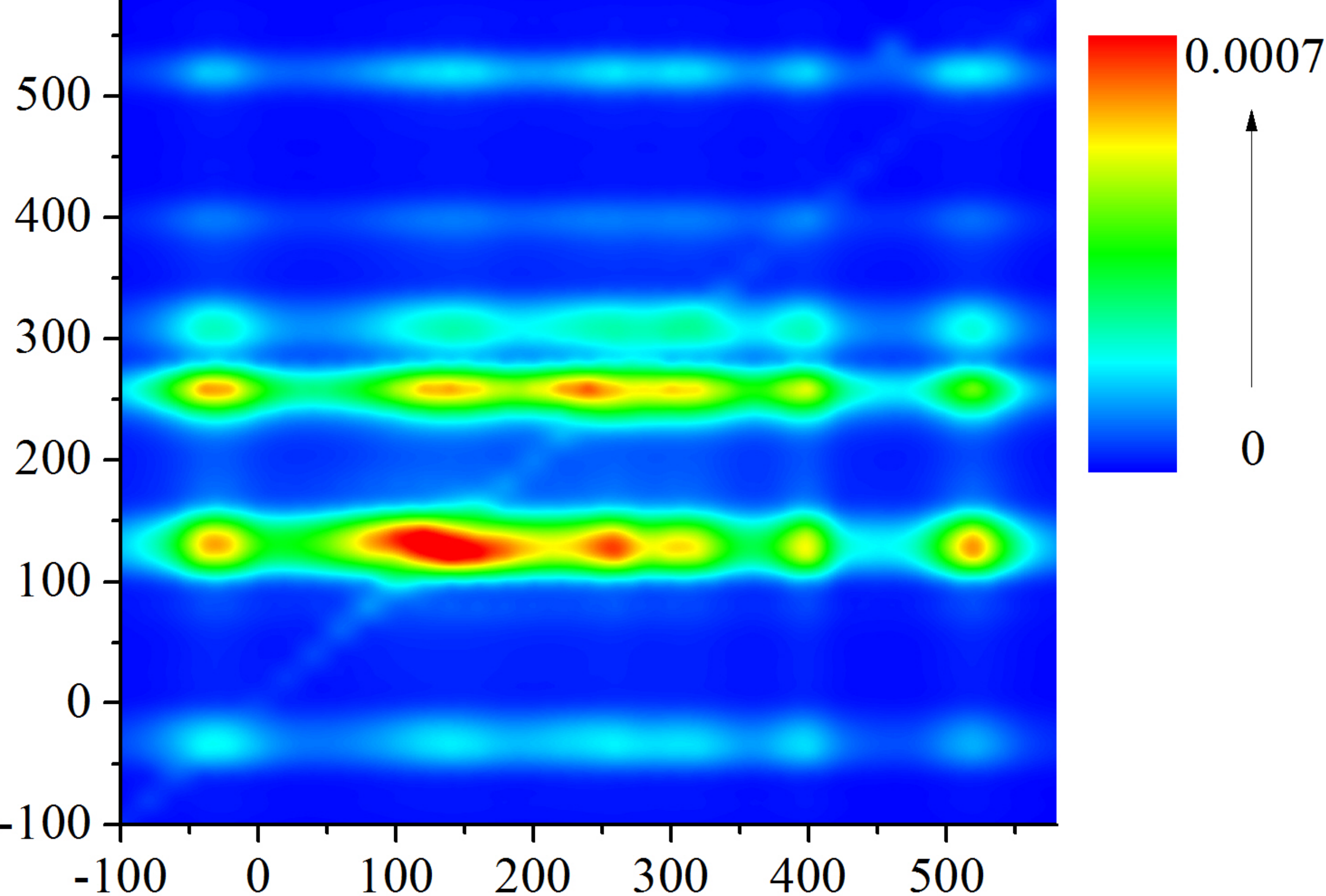}}
\caption{
Stationary mean photon number of the cavity mode as a function
of the cavity ($Q=10^3$) resonance frequency $\omega_c$ (vertical
axis) and the pump laser frequency $\omega_l$ (horizontal),
both in units of cm$^{-1}$ - shifted by 12195 cm$^{-1}$ - for dephasing rate $\gamma=10\ cm^{-1}$.}\label{fig14}
\end{figure}
\begin{figure}[t]
\centerline{\includegraphics[width=.48\textwidth]{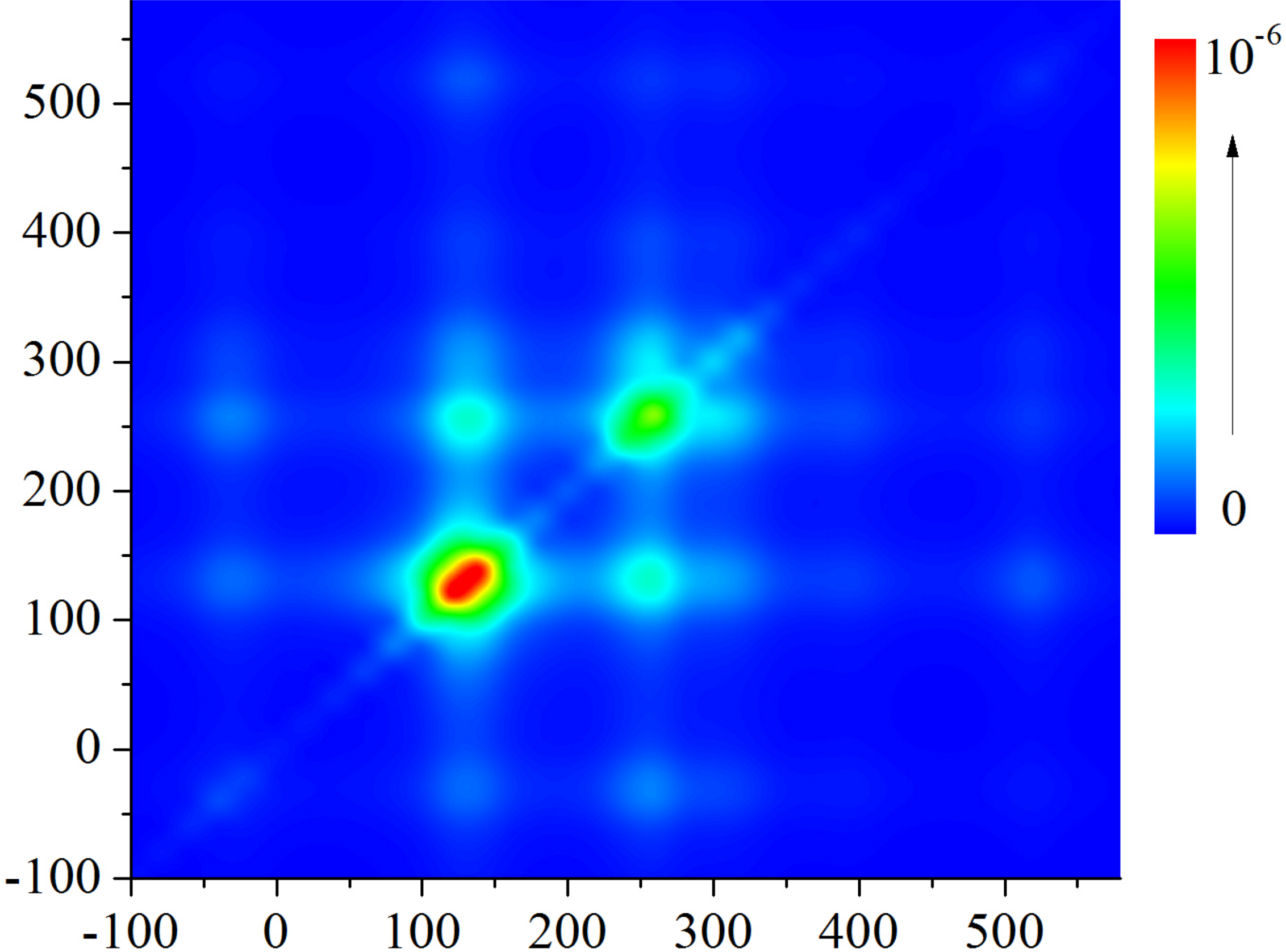}}
\caption{
Stationary mean photon number of the cavity mode as a function
of the cavity ($Q=10^4$) resonance frequency $\omega_c$ (vertical
axis) and the pump laser frequency $\omega_l$ (horizontal),
both in units of cm$^{-1}$ - shifted by 12195 cm$^{-1}$ - in the case of dephasing rate $\gamma=50\ cm^{-1}$, weaker pump field $E=0.01$~cm$^{-1}/D$, and weaker cavity coupling rate $g = 0.01$~cm$^{-1}/D$.}\label{fig18}
\end{figure}
\begin{figure}[t]
\centerline{\includegraphics[width=.48\textwidth]{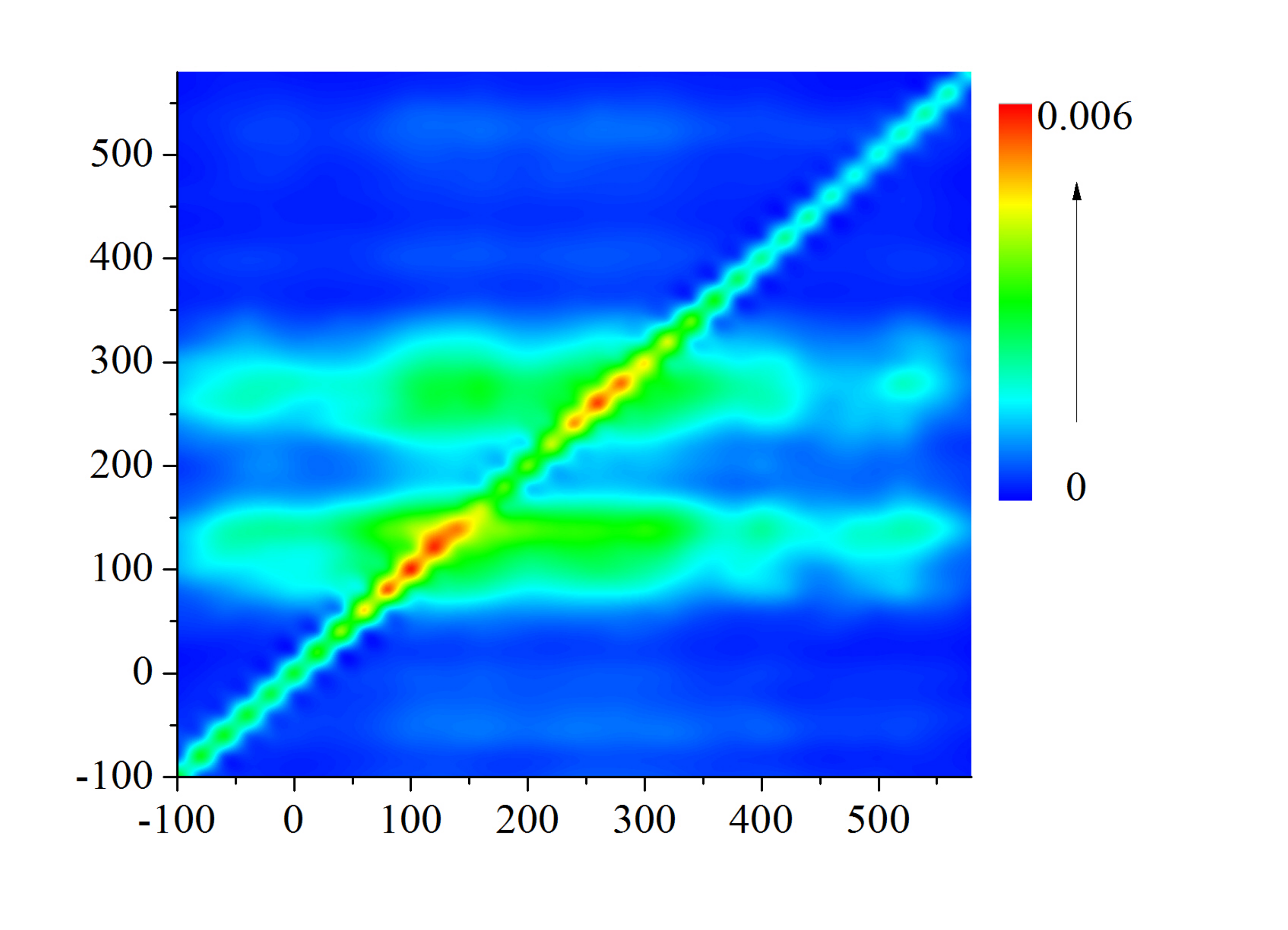}}
\caption{
Stationary mean photon number of the cavity mode as a function
of the cavity ($Q=10^4$) resonance frequency $\omega_c$ (vertical
axis) and the pump laser frequency $\omega_l$ (horizontal),
both in units of cm$^{-1}$ - shifted by 12195 cm$^{-1}$ - in presence of an inhomogeneous broadening,
induced by a static disorder in the site energies, set to $50$ cm$^{-1}$, in the case of dephasing rate $\gamma=10\ cm^{-1}$.}\label{fig17}
\end{figure}
\section{Extracting additional information}
In this section we investigate other important features of the LHC dynamics that we can extract measuring statistical properties of the outcoming light (probe) mode. The stationary mean photon number of the outcoming light mode is shown for different values of dephasing rate $\gamma$ (in Fig. \ref{fig13}), also for a lower-quality ($Q=10^3$) cavity (in Fig. \ref{fig14}), and for a weaker (state-of-the-art) cavity coupling rate (in Fig. \ref{fig18}) according to the present stage of technological advances. The behaviour of the anti-crossing effect withing a diagonal peak as a function of the laser field coupling can be observed in Fig. \ref{fig7}.
\subsection{Inhomogeneous broadening: static disorder}
Here, we show the effect of static disorder, i.e. disorder in the site energies, on the mean photon number 2D spectra, averaged over full random orientation disorder as well (as done in the rest of the manuscript) - see Fig. \ref{fig17} for a disorder of $50$ cm$^{-1}$. As observed in 2D spectroscopy, it lead to the presence of inhomogeneous broadening for the diagonal peaks.
\subsection{Second-order correlation function}
\begin{figure}[t]
\centerline{\includegraphics[width=.45\textwidth]{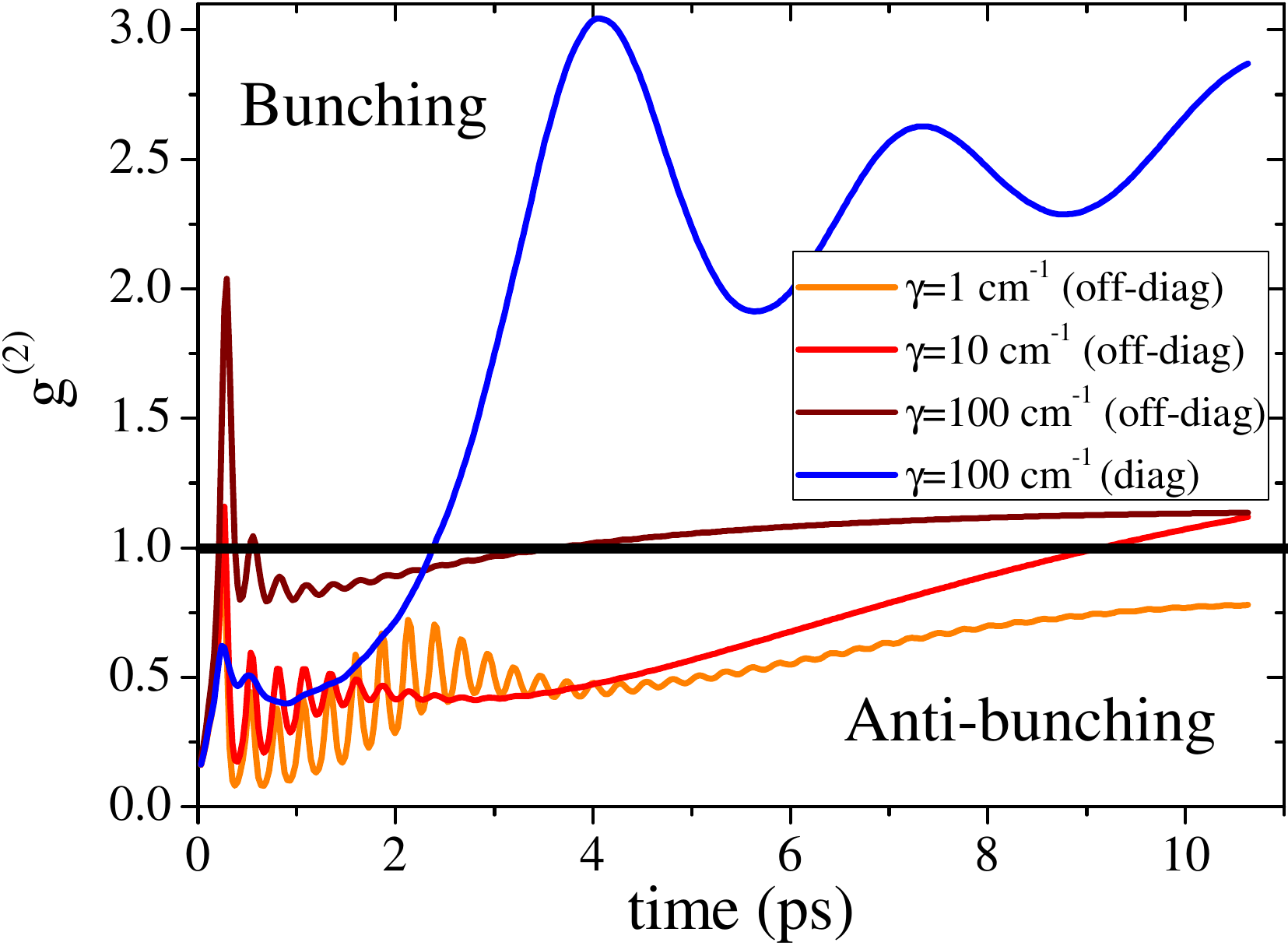}}
\caption{Second-order coherence function at time delay zero $g^{(2)}(0)$ of
the cavity ($Q=10^4$) mode versus time ($\mathrm{ps}$) for
different values of dephasing rates and for diagonal and off-diagonal peaks,
in the case of dephasing rate $\gamma=10\ cm^{-1}$.}\label{fig9}
\end{figure}
\begin{figure}[t]
\centerline{\includegraphics[width=.4\textwidth]{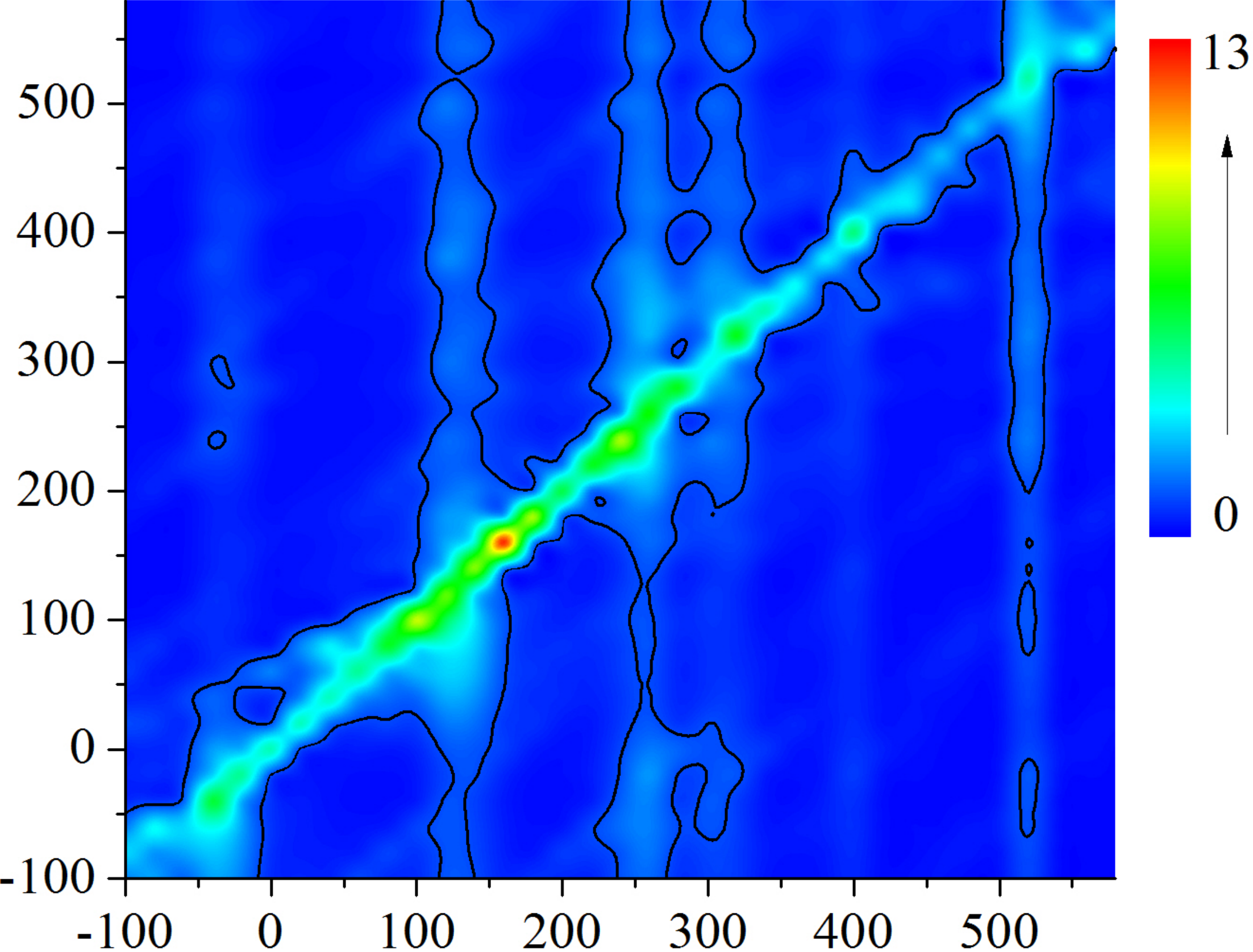}}
\centerline{\includegraphics[width=.4\textwidth]{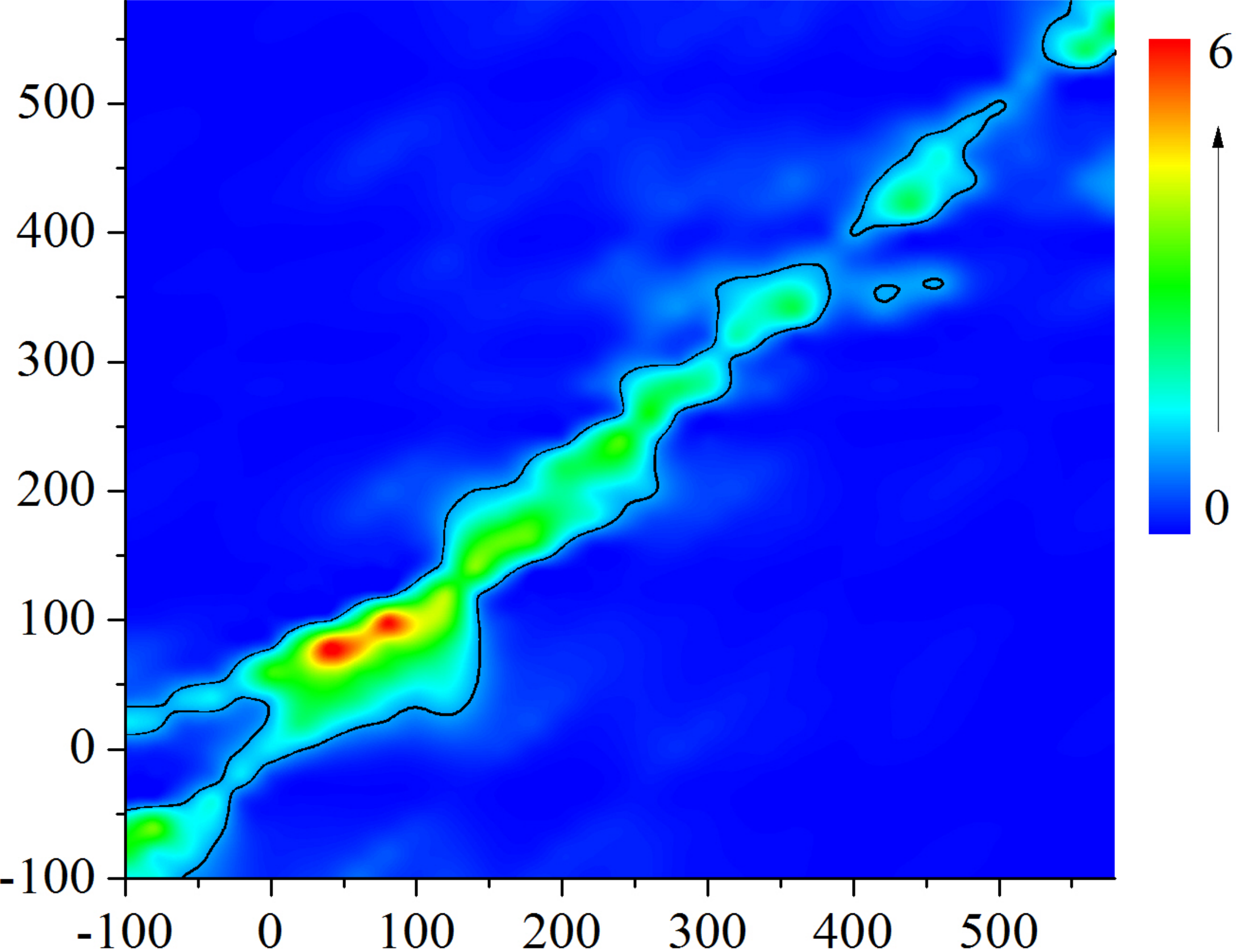}}
\centerline{\includegraphics[width=.4\textwidth]{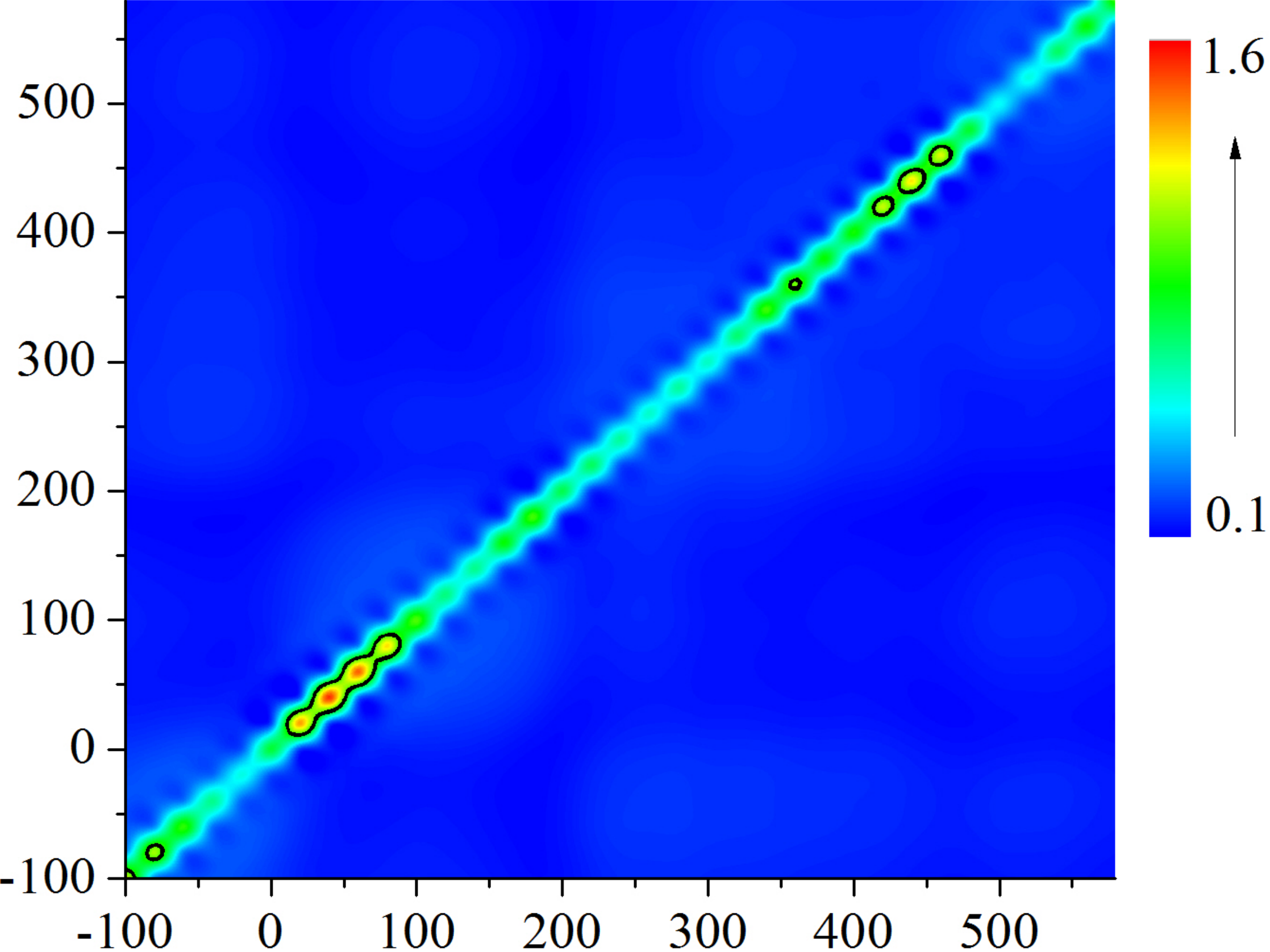}}
\caption{
Second-order coherence function at time delay zero $g^{(2)}(0)$ of
the stationary cavity ($g = 0.1$~cm$^{-1}/D$ and $Q=10^4$ (Top) and $Q=10^3$ (Middle); Bottom: $g = 0.01$~cm$^{-1}/D$,
$E=0.01$~cm$^{-1}/D$, and $Q=10^4$) as a function
of the cavity resonance frequency $\omega_c$ (vertical
axis) and the pump laser frequency $\omega_l$ (horizontal),
both in units of cm$^{-1}$ - shifted by 12195 cm$^{-1}$ - for dephasing rate $\gamma=10\ cm^{-1}$.
The black lines refer to $g^{(2)}(0)=1$.}\label{fig15}
\end{figure}
\begin{figure}[t]
\centerline{\includegraphics[width=.45\textwidth]{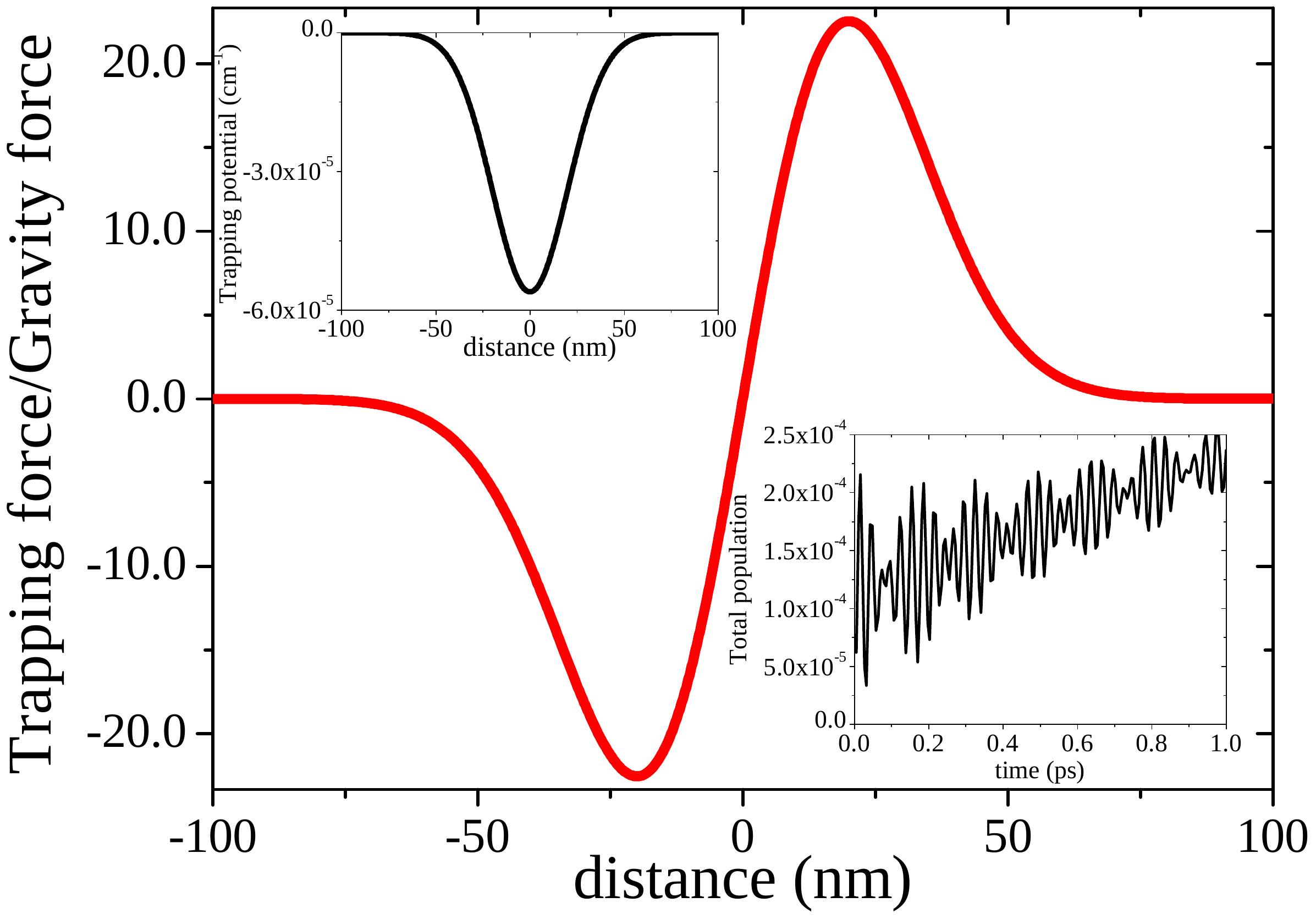}}
\caption{Trapping force (in units of gravity force) vs. position (nm) for a single FMO complex, when subject to a constant laser field with a spatial Gaussian profile peaked in the center of the cavity,
$E_0(x)=\bar{E}_0 e^{-x^2/(2 \lambda)}$ with $\bar{E}_0=5 \ D^{-1} \ cm^{-1} \sim 30 \cdot 10^6 \ J/(C \ m)$, carrier frequency set to $\omega_l=-5000 \ \text{cm}^{-1}$ and polarization orientation given by $\theta \sim 1.75$, $\phi \sim 2$. Top inset: trapping potential (in $\text{cm}^{-1}$) as a function of the position (in $\mathrm{nm}$). Bottom Inset: in the presence of such laser irradiation (even applied for about $1 \ \mathrm{ps}$), the amount of total excitation in the FMO system is essentially vanishing, i.e. smaller than $3 \cdot 10^{-4}$.}\label{fig8}
\end{figure}
Here, we analyze the time evolution of the second-order coherence
function at time delay zero $g^{(2)}(0)$ of the stationary cavity
mode (defined in the main text) for diagonal and off-diagonal
peaks -- see Fig. \ref{fig9}. Although the time scale, we consider
here, is not feasible to be investigated with the current photon
detector technology, however it could be in a near-future or even
nowadays with other natural or artificial light-harvesting systems
where this behavior takes place in a longer time due to different
system parameters. Furthermore, in Figs. \ref{fig15} we show a
full-range contour-plot for the second-order coherence function at
time delay zero $g^{(2)}(0)$ of the stationary cavity mode as a
function of the cavity and pump frequencies, for different cavity
parameters. We observe observe a formation of a quantum photon
state ($g^{(2)}(0) < 1$) almost everywhere, except along the
diagonal line where coherent and thermal states ($g^{(2)}(0) \geq
1$) can appear.
\section{Trapping LHCs inside the cavity}
In our analysis it is important to trap the biological sample inside the cavity. Although several techniques are know in this respect,
in this section we show, by a simple and elementary argument, the feasibility of such process by using an external electric field.
In particular, let us apply an additional constant electric field $E_0$, polarized along the axis $\vec{e}$ with carrier frequency $\omega_l$,
interacting with the FMO complex as the pump laser field. The corresponding interaction energy is
\begin{equation}
\Delta= \sum_{i=1}^{7} \frac{|\vec{\mu}_i \centerdot \vec{e} \ E_0 |^2}{\omega_l-\omega_i} \; .
\end{equation}
and shows a minimum by varying the orientation of the polarization axis $\vec{e}$
in terms of the two angles $\theta$ and $\phi$ with respect the site-1 dipole moment,
i.e., for $\theta \sim 1.75$ and $\phi \sim 2$. In particular,
we consider a spatial profile for the electric field given by a Gaussian function peaked in the center of the cavity as
$E_0(x)=\bar{E}_0 e^{-x^2/(2 \lambda)}$, where $\lambda=800 \ \mathrm{nm}$. In Fig. \ref{fig8}, we show
the trapping potential $\Delta(x)$ and the corresponding trapping force as a function of the position $x$, in the case
of $\theta \sim 1.75$, $\phi \sim 2$, $\omega_l=-5000 \ \text{cm}^{-1}$, and $\bar{E}_0=5 \ D^{-1} \ cm^{-1} \sim 30 \cdot 10^6 \ J/(C \ m)$.
The trapping force is about $20$ times larger than the gravitational force to which a single FMO is subjected, assuming that it has a mass of around
$80 \ \text{kDa} \sim 15 \ 10^{-23} \text{Kg}$ (including the protein scaffolding) \cite{FMO}. In this regime, the amount of electronic excitation in the FMO complex is
very small -- see Inset in Fig. \ref{fig8}. Moreover, we have found that the trapping force can be even several order of magnitudes larger than the gravitational one, allowing a very tiny amount of excitation in the FMO system induced by the presence of the laser irradiation. Therefore, following this simple argument, it seems that it is possible to trap a sample of FMO complexes inside the cavity, without perturbing the system, i.e. leaving it in its ground state, and then experimentally apply a driving laser field as described above.
\section{Entangling bio-samples in separate cavities}
\begin{figure}[t]
\centerline{\includegraphics[width=.4\textwidth]{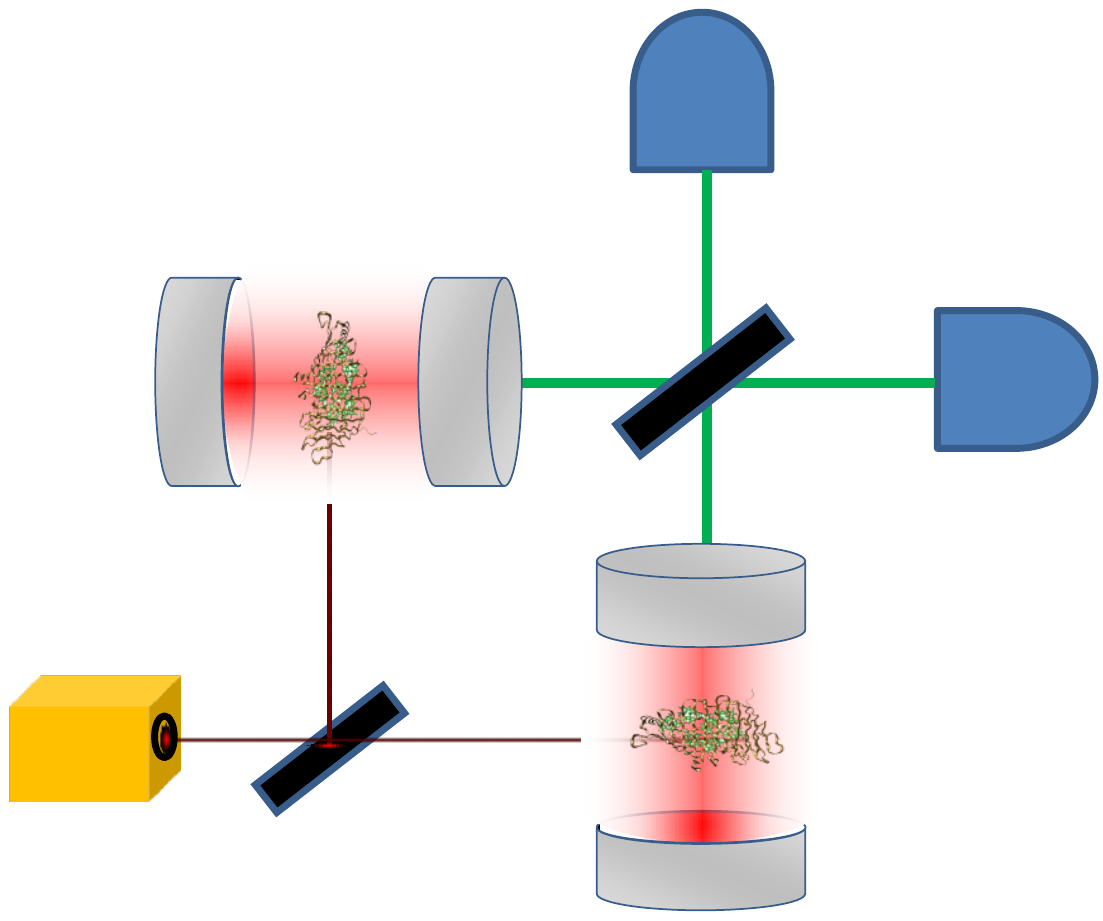}}
\caption{We consider a setup of two equal spatially separate
optical cavities in which two LHC samples are confined, while
subjected to two similar pump laser fields. Two auxiliary
photodectors are also introduced to detect photons, leaking out of
the cavities, after being mixed on a $50/50$
beam-splitter.}\label{fig5}
\end{figure}

\begin{figure}[t]
\centerline{\includegraphics[width=.4\textwidth]{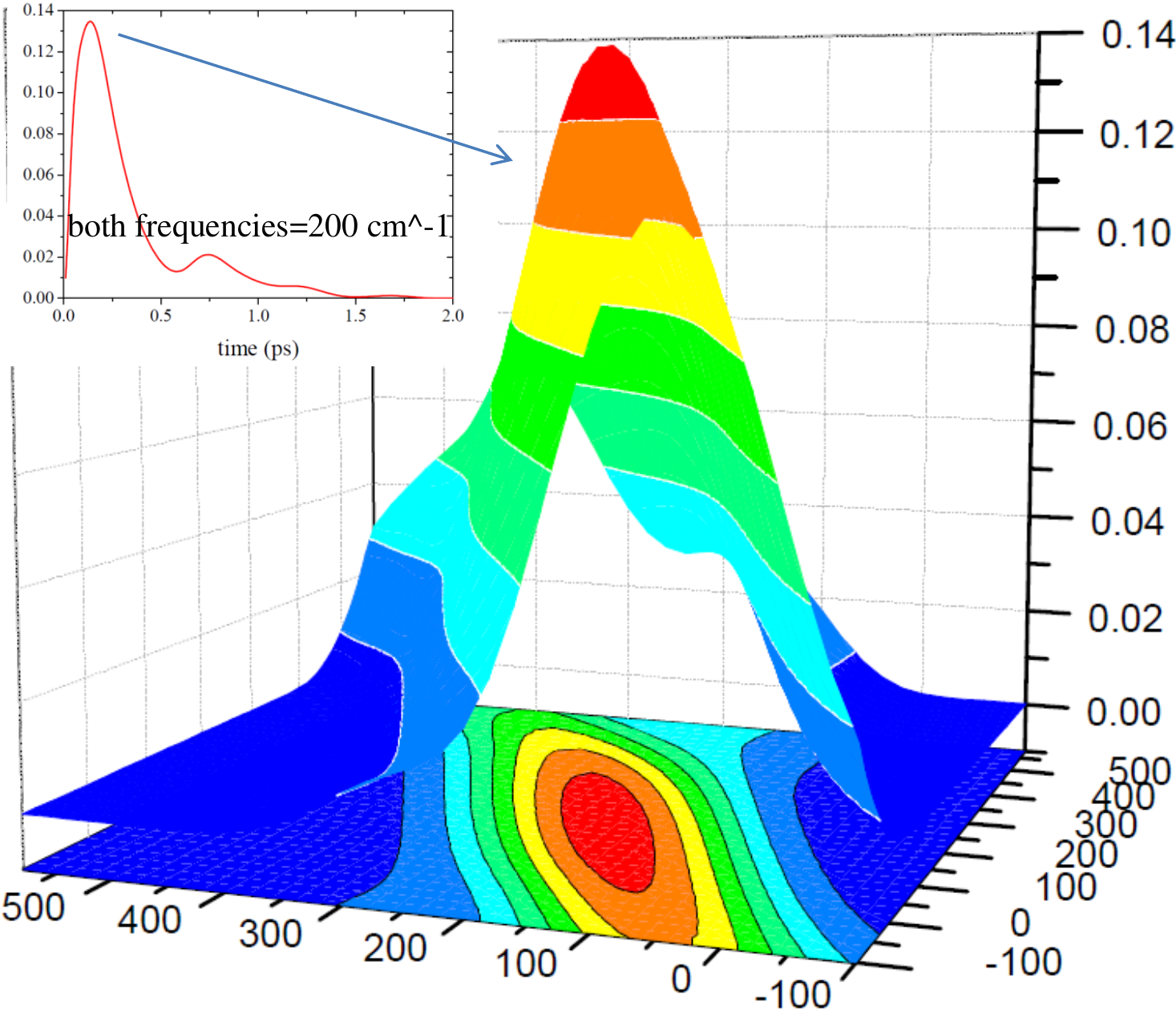}}
\caption{Amount of quantum correlations (measured by logarithmic
negativity) between the two FMO samples trapped in two spatially
separate cavities, after $\sim 140 \mathrm{fs}$, in the case of dephasing rate $\gamma=10\ cm^{-1}$.
Inset: Entanglement vs. time ($\mathrm{ps}$) for $\omega_c=\omega_l=200 \
cm^{-1}$.}\label{fig6}
\end{figure}

Finally, we propose a possible scheme (in Fig. \ref{fig5}) for a probabilistic creation
of quantum correlations, i.e. entanglement, between two FMO
samples confined in spatially separate cavities, following Ref.
\cite{prl2003} where such protocol was introduced for trapped
ions. In other terms, after applying the pump laser fields to the
two separate samples, the outcoming photons from the cavities are
mixed on a $50/50$ beam-splitter and one applies a projection of
the two-photon state into a Bell state of the form $1/\sqrt{2} (
|\text{click}, \text{no click}\rangle + |\text{no click},
\text{click}\rangle)$. Then, after the projection, the composite
system of the two LHC samples has a chance to be left in an
entangled state with some finite probability. Therefore, we
measure the entanglement (by log-negativity) between the two
samples, after the projection for the two-cavity mode into the
Bell state. This quantity is shown in Fig. \ref{fig6}. We find
that one can probabilistically entangle the two biological samples (entangled polaritons)
with a non-vanishing probability. This scheme may, for instance, allow us to extract more information
on the quantum performance of these biological molecules, and perhaps pave the way for
new tests about the foundation of quantum mechanics in terms of biological systems
where quantum features seem to play an important role.


\begin{thebibliography}{99}

\bibitem{prok}
V.I. Prokhorenko, A.R. Holzwarth, F.R. Nowak, and T.J. Aartsma,
\textit{J. Phys. Chem. B} \textbf{106}, 9923 (2002).

\bibitem{fleming07a} G.S. Engel, T.R. Calhoun, E.L. Read, T.-K. Ahn, T. Manal, Y.-C. Cheng, R.E.
Blankenship, and G.R. Fleming, \textit{Nature} \textbf{446}, 782 (2007);
H. Lee, Y-C. Cheng, and G.R. Fleming, \textit{Science} \textbf{316}, 1462 (2007).

\bibitem{Engel2010}
G. Panitchayangkoon, D. Hayes, K.A. Fransted, J.R. Caram, E. Harel, J. Wen, R.E. Blankenship,
and G.S. Engel, \textit{Proc. Natl. Acad. Sci.} \textbf{107}, 12766 (2010).

\bibitem{collini}
E. Collini, C.Y. Wong, K.E. Wilk, P.M.G. Curmi, P. Brumer, and G.D. Scholes,
\textit{Nature} \textbf{463}, 644–648 (2010).

\bibitem{marangos}
I.P. Mercer, Y.C. El-Taha, N. Kajumba, J.P. Marangos, J.W.G. Tisch, M. Gabrielsen, R.J. Cogdell, E. Springate, and E. Turcu,
\textit{Phys. Rev. Lett.} \textbf{102}, 057402 (2009).

\bibitem{Foerster}
T. F\"{o}rster, Delocalized excitation and excitation transfer., in Modern Quantum Chemistry, Sinanoglu, O. (ed.), pages 93-137 (New York and London: Academic Press, 1965).

\bibitem{Aspuru08} M. Mohseni, P. Rebentrost, S. Lloyd, and A. Aspuru-Guzik,
 \textit{J. Chem. Phys.} {\bf 129}, 174106 (2008).

\bibitem{Plenio08}
M.B. Plenio and S.F. Huelga, \textit{New J. Phys.} {\bf 10}, 113019 (2008).

\bibitem{patrick}
P. Rebentrost, M. Mohseni, I. Kassal, S. Lloyd, and A. Aspuru-Guzik,
\textit{New J. Phys.} {\bf 11}, 033003 (2009).

\bibitem{castro08}
A. Olaya-Castro, C.F. Lee, F. Fassioli Olsen, and N.F. Johnson, \textit{Phys. Rev.} \textbf{B} 78, 085115 (2008).

\bibitem{ccdhp09}
F. Caruso, A.W. Chin, A. Datta, S.F. Huelga, and M.B. Plenio, \textit{J. Chem. Phys.} \textbf{131}, 105106 (2009);
\textit{Phys. Rev. A} \textbf{81}, 062346 (2010).

\bibitem{cdchp10}
 A.W. Chin, A. Datta, F. Caruso, S.F. Huelga, and M.B. Plenio, \textit{New J. Phys.} \textbf{12}, 065002 (2010).

\bibitem{chp}
F. Caruso, S.F. Huelga, and M.B. Plenio, \textit{Phys. Rev. Lett.} \textbf{105}, 190501 (2010).

\bibitem{alan1}
S. Shim, P. Rebentrost, S. Valleau, and A. Aspuru-Guzik, Eprint arXiv:1104.2943 (2011).

\bibitem{alan2}
J. Yuen-Zhou, M. Mohseni, J.J. Krich, and A. Aspuru-Guzik, \textit{Proc. Natl. Acad. Sci. USA}, in press (2011), Eprint arXiv:1006.4866.

\bibitem{alan3}
P. Rebentrost and A. Aspuru-Guzik, \textit{J. Chem. Phys.} \textbf{134}, 101103 (2011).

\bibitem{alan4}
J. Zhu, S. Kais, P. Rebentrost, and A. Aspuru-Guzik, \textit{J. Phys. Chem. B} \textbf{115}, 1531-1537 (2011).

\bibitem{Mabuchi02}
H. Mabuchi and A. C. Doherty, \textit{Science} {\bf 298}, 1372
(2002).

\bibitem{Finley11}
F. P. Laussy, E. del Valle, M. Schrapp, A. Laucht, J. J.
Finley, Eprint arXiv:1104.3564 (2011).

\bibitem{Khitrova2006}
G. Khitrova, H.M. Gibbs, M. Kira, S.W. Koch, and A. Scherer, \textit{Nat. Physics} \textbf{2}, 81-90 (2006).

\bibitem{vuckovic}
A. Majumdar, A. Papageorge, E.D. Kim, M. Bajscy, H. Kim, P. Petro,
and Jelena Vu\v{c}kovi\'{c}, Eprint arxiv:1103.2716 (2011); A. Majumdar, M. Bajscy, and J. Vu\v{c}kovi\'{c}, Eprint arxiv:1106.1926 (2011).

\bibitem{Murr11}
A. Ourjoumtsev, A. Kubanek, M. Koch, C. Sames, P. W. H. Pinkse, G.
Rempe, and K. Murr, \textit{Nature} {\bf 474}, 623 (2011).

\bibitem{hartmann}
M.J. Hartmann, F.G.S.L. Brand\~{a}o, and M.B. Plenio, \textit{Nat. Phys.} \textbf{2}, 849 (2006).

\bibitem{angelakis}
D.G. Angelakis, M.F. Santos, and S. Bose, \textit{Phys. Rev. A} \textbf{76}, 031805(R) (2007).

\bibitem{greentree}
A.D. Greentree, C. Tahan, J.H. Cole, and L.C.L. Hollenberg, \textit{Nat. Phys.} \textbf{2}, 856 (2006).

\bibitem{Kasprzak2006}
J. Kasprzak \textit{et al.}, \textit{Nature} \textbf{443}, 409 (2006).

\bibitem{Deng2010}
H. Deng and Y. Yamamoto, \textit{Rev. Mod. Phys.} \textbf{82}, 1489 (2010).

\bibitem{Butte2011}
R. Butt\'{e} and N. Grandjean, \textit{Semicond. Sci. Tech.}
\textbf{26}, 014030 (2011).

\bibitem{Kena-Cohen2010}
S. K\'{e}na-Cohen and S.R. Forrest, \textit{Nat. Photonics} \textbf{4}, 371 (2010).

\bibitem{mabuchi}
T. McGarvey, A. Conjusteau and H. Mabuchi, \textit{Optics Express} {\bf 14} 10449 (2006).

\bibitem{rempe}
M. Koch, C. Sames, M. Balbach, H. Chibani, A. Kubanek, K. Murr,
T. Wilk, and G. Rempe, \textit{Phys. Rev. Lett.} {\bf 107}, 023601 (2011).

\bibitem{CNRS_PRL2010}
R. Gehr, J. Volz, G. Dubois, T. Steinmetz, Y. Colombe,
B.L. Lev, R. Long, J. Est\'eve, and J. Reichel, \textit{Phys. Rev. Lett.} {\bf 104}, 203602 (2010).

\bibitem{FP_cavity_NJP2010}
D. Hunger, T. Steinmetz, Y. Colombe, C. Deutsch, T. W. H\"ansch,
and J. Reichel, {\it New J. Physics} {\bf 12}, 065038 (2010).

\bibitem{note}
Very recently, it has been discovered that the FMO subunit has $8$
pigments. However, this should not affect the analysis we propose
here for a generic light-harvesting system.

\bibitem{Fleming2005}
M. Cho, H.M. Vaswani, T. Brixner, J. Stenger, and G.R. Fleming, {\it J. Phys. Chem. B} {\bf 109}, 10542 (2005).

\bibitem{adolphs}
J. Adolphs and T. Renger, \textit{Biophys. J.} \textbf{91}, 2778, (2006).

\bibitem{haken}
H. Haken and G. Strobl, \textit{Z. Phys.} \textbf{262}, 135 (1973).

\bibitem{aki}
A. Ishizaki and G.R. Fleming, \textit{Proc. Natl Acad. Sci. USA} \textbf{106}, 7255 (2009).

\bibitem{patrickNJP}
P. Rebentrost, R. Chakraborty, A. Aspuru-Guzik, \textit{J. Chem. Phys.} \textbf{131}, 184102 (2009).

\bibitem{prior2010}
J. Prior, A.W. Chin, S.F. Huelga, and M.B. Plenio, \textit{Phys. Rev. Lett.} \textbf{105}, 050404 (2010).

\bibitem{thorwart2009}
M. Thorwart, J. Eckel, J.H. Reina, P. Nalbach, and S. Weiss, \textit{Chem. Phys. Lett.} \textbf{478}, 234 (2009).

\bibitem{hayes2010}
D. Hayes, G. Panitchayangkoon, K.A. Fransted, J.R. Caram, J. Wen, K.F. Freed, and G.S. Engel, \textit{New J. Phys.} \textbf{12}, 065042 (2010).

\bibitem{MandelWolf}
L. Mandel and E. Wolf, {\it Optical Coherence and Quantum Optics} (Cambridge University Press, Cambridge, 1995).

\bibitem{plenio05}
M.B. Plenio, \textit{Phys. Rev. Lett.} \textbf{95}, 090503 (2005).

\end{thebibliography}

\begin{thebibliography}{99}

\bibitem{Renger2008}
J. Adolphs, F. M\"uh, M. El-Amine Madjet and Th. Renger, Photosynth. Res. \textbf{95},
197 (2008).

\bibitem{adolphs06} J. Adolphs and T. Renger, \textit{Biophys. J.} \textbf{91}, 2778, (2006).

\bibitem{PDB}
Structure from PDB: 3EOJ; D.E. Tronrud. J. Wen, L. Gay, and R.E. Blankenship,  \textit{Photosynth. Res.} \textbf{100}, 79-87 (2009).

\bibitem{FMO}
M.T.W. Milder, B. Br\"{u}ggemann, v.R. Grondelle, J.L. Herek,
\textit{Photosynth. Res.} \textbf{194}, 257-274 (2010).

\bibitem{prl2003}
D.E. Browne, M.B. Plenio, and S.F. Huelga, \textit{Phys. Rev. Lett.} \textbf{91}, 067901 (2003).

\end{thebibliography}
\end{document}